\documentclass[12pt]{article}

\usepackage{graphics}
\usepackage{float}
\usepackage{epsfig}
\usepackage{epstopdf}
\usepackage{setspace}
\title{Interacting modified holographic Ricci dark energy model and statefinder diagnosis 
in flat universe}

\author{Praseetha P and Titus K Mathew \\
Department of Physics, \\ Cochin University of Science and Technology, \\ Kochi-682022,
India \\
E-mail:praseethapankunni@gmail.com and titus@cusat.ac.in}
\date{}
\begin{document}

\maketitle

\abstract{
In this work we have considered the modified holographic Ricci dark energy interacting with dark matter through a non-gravitational 
coupling. We took three phenomenological forms for the interaction term $Q$ in the model, where in general $Q$ is proportional to the 
Hubble parameter and densities of the dark sectors, $\rho_{de}+\rho_m, \, \rho_m$ and $\rho_{de}$ respectively. We have obtained 
analytical solutions for the three interacting models, and studied the evolutions of equations of state parameter, deceleration 
parameter. The results are compared with the observationally constrained values for the best parameters of the model. We have also done the statefinder 
analysis of the model to discriminate the model from other standard models. In general we have shown that the model is showing a de Sitter 
type behavior in the far future of the evolution of the universe.} 



%
%
\section{Introduction}
\label{intro}
The recent observational advance in cosmology have shown that, the expansion rate of the universe is accelerating \cite{Perl,Ries}. 
This discovery was announced in 1998 based on the data on Supernovae Type IA. A series of subsequent experiments regarding the Cosmic
Microwave Background (CMB) radiations, Large Scale Structure (LSS) and other measurements have all confirmed the above claim on
accelerating expansion of the universe. Theoretical analysis which explaining this recent acceleration, predicts that, the major 
component of the universe is dark energy, an exotic cosmic component with negative pressure. However the nature of dark energy is still
enigmatic, so that it's nature is one of the biggest challenges in cosmology. Cosmological constant $\Lambda$ is an important candidate 
for dark energy and it provides a good explanation for the current acceleration. But cosmological constant faces the severe draw backs
such that, the theoretical value of $\Lambda$ is many orders of magnitude larger than the current observational value \cite{wein1,sahni1,Cai1,Li1}
and it is not able to explain, why densities of dark energy and dark matter are of same order at present while they evolve in rather different
ways. So as an alternative, dynamical dark energy models have been proposed and analyzed in recent literature. Among these holographic 
dark energy models \cite{Xang1,Huang1,Xang2,Xang3} have got much recent attention, because it originates form the holographic principle 
of quantum gravity \cite{Skind1}. Ac   cording to this principle, the vacuum energy density can be bounded as 
$\rho_{vac} L^3 \leq M_{pl}^2 L,$ \cite{Cohen1}, where $\rho_{vac}$ is the vacuum energy density and $M_{pl}$ is the reduced plank mass.
This bound implies that, the total energy inside a region of size $L$, should not exceed the mass of a black hole of the same size. From
effective quantum field theory, an effective IR cut of can saturate the length scale, so that the dark energy density (the vacuum energy 
density) can be written as $\rho_{de} = 3 c^2 M_{pl} L^{-2}$ \cite{Li2}, where $c$ is a dimensionless numerical factor. In literature, 
the IR cut-off has been taken as the Hubble horizon \cite{Li2,Hsu1}, particle horizon, event horizon \cite{Li2} and some generalized 
IR cut-off \cite{Gao1,Zhang1,Yang1}. The holographic dark energy models with Hubble horizon and particle horizon as the IR cut-off, 
cannot lead to the current accelerated expansion of the universe. When event horizon is taken as the cut-off, the model is suffered 
from the following problem. Future event horizon is global concept of space-time, while dark energy density is a local quantity. So the 
relation between them will pose challenges to causality. These leads to the introduction of the new holographic dark energy, the 
holographic Ricci dark energy, where the IR cut-off is taken as proportional to the Ricci scalar curvature, $R^{1/2}.$ The holographic 
Ricci dark energy introduced by Granda and Oliveros \cite{Granda1} is in fairly good agreement with the observational data. This model have
the following advantages. The fine tuning problem can be avoided in this model of dark energy. Moreover the presence of event horizon is 
not presumed in this model, so that the causality problem can be avoided and the coincidence problem can also be avoided in this model.
Recently a modified form of holographic Ricci dark energy in interaction with the dark matter was analyzed \cite{Luis1}.

The current observational evidence indicates that around 95$\%$ of matter-energy in the universe is in the dark sector, composed of 
dark matter and dark energy \cite{Perl,Ries,Teg1}. Dark matter is substantially non-baryonic in nature and would responsible for the 
structure formation in the universe. There is an unavoidable degeneracy between dark matter and dark energy existing within the Einstein's
gravity. So there could be a hidden non-gravitational coupling between them. This interestingly leads to develop various ways of testing 
different kinds of interaction in the dark sector. in the present paper we consider the interaction between dark energy and dark matter, 
by considering dark energy density as the modified holographic dark energy. Owing to the lack of mechanism for the microscopic origin of 
the interaction, one has to assume various forms for the interaction phenomenologically. Several forms for the interaction have been put 
forward \cite{Amend1,Xang4,Nijor1,He1,Li3}. The commonly used interaction forms are those in which the interaction is depend linearly on the 
Hubble parameter and the density of dark matter, dark energy or sum of both the densities. In all these works the dark energy taken as 
the holographic dark energy. As we have mentioned, in this work we have considered the modified holographic Ricci dark energy in 
interaction with the dark matter. 

Statefinder parameters, introduced by Sahni et al. \cite{Sahni2} is a sensitive diagnostic tool used to discriminate various dark energy 
models, because the Hubble parameter and the deceleration parameter lone cannot effectively discriminate various dark energy models.
These parameters are defined as,
\begin{equation} \label{eqn:rs}
 r= {1 \over a H^3} {d^3 a \over dt^3} \, \, \, \, , \, \, \, s = {r - \Omega_{total} \over 3(q - \Omega_{total}/2)} 
\end{equation}
where $a$ is the scale factor and $\Omega_{total}$ is the total energy density parameter containing the matter and dark energy. 
The $r-s$ plot can discriminate the various dark energy models, for example, the well known $\Lambda$CDM model is one with,
$r=1$ and $s=0.$ The cosmological behavior of various dark energy models were differentiated using the statefinder parameters 
\cite{Setare1,Malek1}.

\section{Interacting MHRDE model}
\label{sec:1}
The Friedmann equation for the flat universe with FRW metric is,
\begin{equation} \label{eqn:friedmann1}
 3H^2=\rho_m + \rho_{de}
\end{equation}
where $H$ is the Hubble parameter, $\rho_m$ is the dark matter density and $\rho_{de}$ is the dark energy density. We have considered the
flat universe because the inflationary model of the universe predicted a flat universe which has been confirmed by observations that the 
current density parameter corresponds to curvature is $\Omega_k \sim 10^{-3}$ \cite{Komatsu1}.  The modified Holographic Ricci dark 
energy (MHRDE), taking Ricci scalar as the IR cut-off is given as
\begin{equation} \label{eqn:mhrde}
 \rho_{de} = {2 \over \alpha - \beta} \left(2 \dot{H} + \frac{3}{2} \alpha H^2 \right) 
\end{equation}
where $\dot{H}$ is the derivative of the Hubble parameter with respect to cosmic time, $\alpha$ and $\beta$ are constants, the model 
parameters. This model was studied in the non-interacting case in reference \cite{tkm1}, and Chimento et. al. have analyzed this this 
type of dark energy in interaction with dark matter as Chaplygin gas \cite{Luis1,Luis2}. The interaction between MHRDE and dark matter 
can be included through the continuity equations,
\begin{equation} \label{eqn:continuity1}
 \dot{\rho}_{de} + 3 H \left( \rho_{de} + p_{de} \right)= -Q
\end{equation}
\begin{equation} \label{eqn:continuity2}
 \dot{\rho}_m + 3 H \rho_m = Q
\end{equation}
Where $p_{de}$ is the pressure density of dark energy, $Q$ is the interaction term, over-dot representing derivative with respect to
time and dark matter is assumed to be pressure less. Since there 
is no conclusive theory for the microscopic origin of the interaction, one has to assume the form of $Q$ phenomenologically. The 
interaction term must be a function of a quantity with dimension inverse of time and thus $Q$ can take forms \cite{Fu1,Chatto1} such as 
$Q=3bH (\rho_{de}+\rho_m)$, $Q=3bH\rho_m$ and $Q=3bH\rho_{de}.$ By convention, $b>0$ means energy is transferring from dark energy to 
cold dark matter. For convenience we will abbreviate the three interacting models as: IMHRDE1 corresponds to $Q=3bH (\rho_{de}+\rho_m)$, 
IMHRDE2 corresponds to $Q=3bH\rho_m$ and IMHRDE3 corresponds to $Q=3bH\rho_{de}.$

\subsection{Interacting model with $Q=3bH(\rho_{de}+\rho_m)$-IMHRDE1}
\label{sec:2} 

In this section we are analyzing the interaction of the MHRDE with cold dark matter, with interaction given as $Q=3bH(\rho_{de}+\rho_m).$
Substituting the MHRDE density equation (\ref{eqn:mhrde}) in the Friedmann equation (\ref{eqn:friedmann1}), we get
\begin{equation}
 h^2 = {\rho_m \over 3H_0^2} + {2 \over 3\Delta} \left(\frac{1}{2} {dh^2 \over dx} + \frac{3\alpha}{2} h^2 \right)
 \end{equation}
where $h = H/H_0$, $H_0$ is the present value of the Hubble parameter, $\Delta=\alpha-\beta$ and the variable $x=log a$ with $a$ as the 
scale factor of the universe. Differentiate this equation once more and substituting $\dot{\rho}_m$ form the continuity equation, leads 
to
\begin{equation} \label{eqn:diff1}
 {d^2h^2 \over dx^2} + 3 \left(1+\beta \right) {dh^2 \over dx} + 9 \left(\beta + b \Delta \right) h^2 = 0
\end{equation}
The solution of the above second order differential equation is obtained as,
\begin{equation} \label{eqn:h2}
 h^2 = c_1 e^{\frac{3}{2} m_1 x} + c_2 e^{\frac{3}{2} m_2 x}
\end{equation}
where
\begin{equation}
 m_{1,2} = -1 -\beta \mp \sqrt{1-4b \alpha - 2\beta + 4 b \beta + \beta^2}.
\end{equation}
The coefficients $c_1$ and $c_2$ are determined using the initial conditions,
\begin{equation}
 h^2|_{x=0} = 1 ,  \, \, \, \, {dh^2 \over dx}|_{x=0} = 3 \Omega_{de0} \Delta - 3 \alpha
\end{equation}
where $\Omega_{de0}$ is the current value of dark energy density and is related to matter density as $\Omega_{de0}=1-\Omega_{m0}$ for the 
flat universe
and $\Omega_{m0}$ is present value of cold dark matter density. From these the coefficients $c_1$ and $c_2$ are found to be,
\begin{equation}
 c_1 = {2\left(\Omega_{de0} \Delta - \alpha \right) - m_2 \over m_1 - m_2} , \, \, \, \, c_2 = 1 - c_1 
\end{equation}
Comparing equation (\ref{eqn:h2}) with the standard Friedmann equation, the dark energy density can be identified as,
\begin{equation} \label{eqn:deomega1}
 \Omega_{de} = c_1 e^{\frac{3}{2} m_1 x} + c_2 e^{\frac{3}{2} m_2 x} - \Omega_{m0} e^{-3x}
\end{equation}
The pressure of the dark energy can then be obtained as
\begin{equation}
 p_{de} = -\Omega_{de} - \frac{1}{3} {d \Omega_{de} \over dx}=-\left[c_1\left(1+\frac{m_1}{2}  \right) e^{\frac{3}{2}m_1 x} + 
c_2\left(1+\frac{m_2}{2}  \right) e^{\frac{3}{2}m_2 x} \right]
\end{equation}
The corresponding equation of state can be obtained using the standard relation,
\begin{equation}
 \omega_{de} = -1 - \frac{1}{3} {d \ln \Omega_{de} \over dx}
\end{equation}
which after using the equation (\ref{eqn:deomega1}), gives,
\begin{equation}
 \omega_{de} = -1- \frac{1}{2} \left({c_1 m_1 e^{\frac{3}{2}m_1 x} + c_2 m_2 e^{\frac{3}{2}m_2 x} + 2 \Omega_{m0} e^{-3x} 
\over c_1  e^{\frac{3}{2}m_1 x} + c_2 e^{\frac{3}{2}m_2 x} -  \Omega_{m0} e^{-3x} } \right)
\end{equation}
If there is no interaction between the dark sectors, i.e. $b=0$, and the contribution form non-relativistic cold dark matter behavior 
($\sim \Omega_{mo}$) term is negligible in the dark energy density, the constants takes the values $m_1=-2, m_2=-2\beta, c_1=0$ and 
$c_2=1.$ Consequently the equation of state parameter become
\begin{equation}
 \omega_{de} = -1 + \beta.
\end{equation}
This is in agreement with the earlier results in the non-interacting case \cite{tkm1}. So in the non-interacting case the equation 
of state can be greater than or less than -1, depending on the value of the the parameter $\beta.$

\begin{figure}[h]
 \includegraphics[scale=0.75]{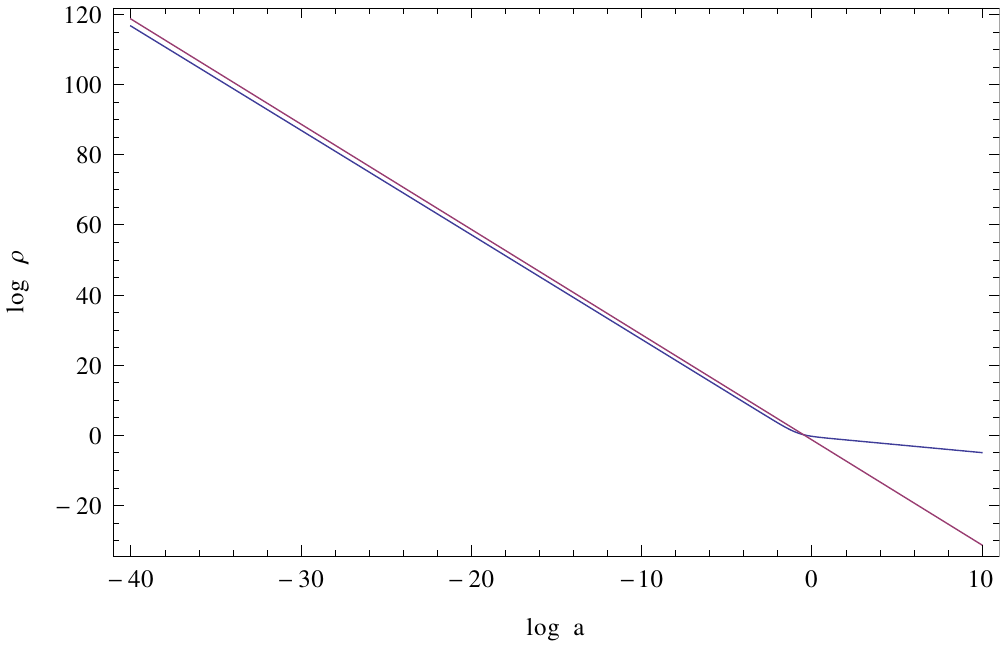}
\caption{Evolution of IMHRDE1 in comparison with the cold dark matter, with parameters $(\alpha, \beta)$ = (1.15,0.15) and 
$b$=0.001. Blue line representing MHRDE and violet line is for cold dark matter.}
\label{fig:evolution1}
\end{figure}

We have analyzed the equation of state parameter for different parameter values. The interaction parameter $b$, is chosen to be
$b=0.001$, because for values greater than this IMHRDE1 does not satisfy the coincidence of matter and dark energy (the 
coincidence problem), for the $(\alpha, \beta)$ parameter sets using by us. The evolution of the IMHRDE1 along with the dark matter is 
shown in the figure \ref{fig:evolution1}.
In studying this co-evolution of dark energy and dark matter, we have neglected the phase transitions, transitions form non-relativistic to relativistic particles at high temperature and also 
new degrees of freedom thus arises. However it is expected that these will not make much changes in the result. The plot shows that, the 
interacting dark energy and the dark matter were comparable with each other in the past universe and MHRDE is dominating at low redshift,
which in effect solves the coincidence problem. We have found that the IMHRDE1 is compatible with the coincidence of matter and dark energy
for all the parameter values of $(\alpha, \beta)$ we used through out our analysis.

\begin{figure}[h]
 \includegraphics[scale=0.75]{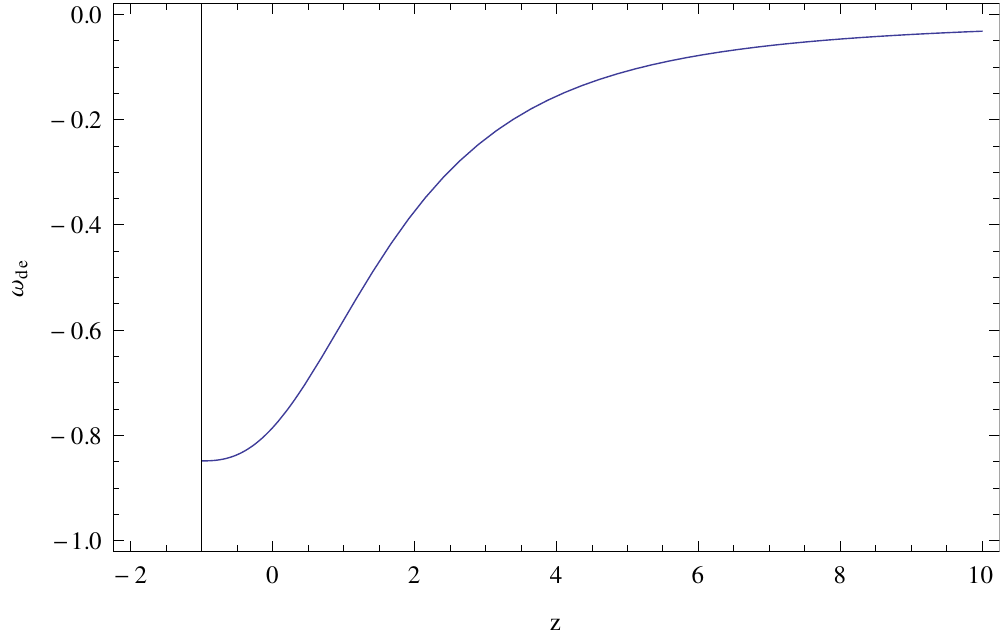}
\caption{Variation of the equation of state parameter $\omega_{de}$ with redshift $z$ for the model parameters 
$\alpha=1.15, \, \, \beta=0.15$ and $b=0.001$}
\label{fig:eos1}
\end{figure}

 We have plotted the equation of state with redshift for the parameters 
$(\alpha, \beta) = (1.15, 0.15)$, for the standard values of $\Omega_{de0}=0.7$, and 
$\Omega_{m0}=0.3.$ 
The plot is shown in the figure \ref{fig:eos1}.
The evolution of the equation of state of IMHRDE1 shows that, in the remote past, at 
large redshift, the equation of state parameter $\omega_{de}$ of the dark energy is very near to zero, in which it 
behaves like cold dark matter. 
\begin{figure}[h]
 \includegraphics[scale=0.85]{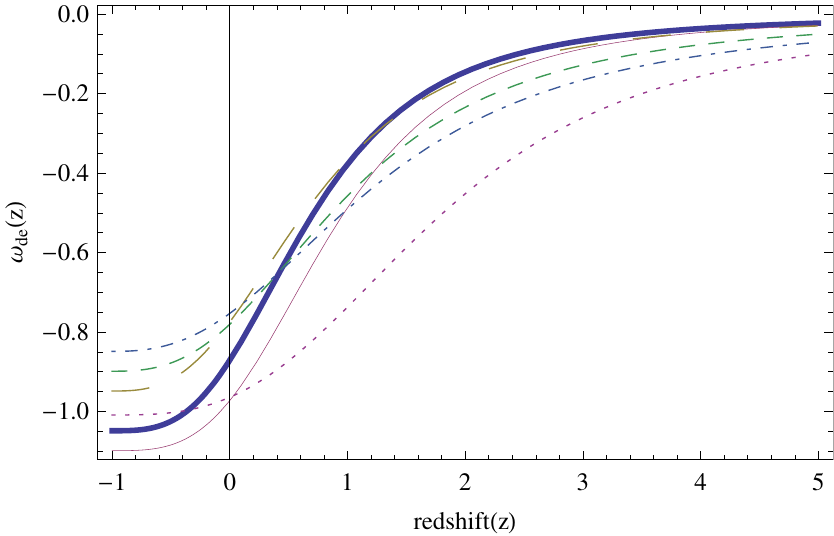}
\caption{Variation of the equation of state parameter $\omega_{de}$ with redshift $z$ for the model parameters 
$(\alpha , \beta)$=(4/3, -0.05)-thick continuous line, (4/3,0.05) large dashed line (1.15,0.15)-dot-dashed line,(1.2,0.1)small dashed line,
 (1.2,-0.1)- thin continuous line and (1.01,-0.01) doted line, with $b$=0.001}
\label{fig:eos2}
\end{figure}
But in the far future, as redshift $z \rightarrow -1,$ equation of state approaches a negative saturation value. From the figure 
\ref{fig:eos1}
the current value of the equation of state is found to be around $\omega_{de} \sim -0.82$ for parameters $(\alpha,\beta)$=(1.15,0.15).
 From figure \ref{fig:eos2} it is clear that for the parameter set (1.2,0.1), $\omega_{de}$=-0.78 and as $z\rightarrow$-1, the equation of 
state parameter shows behavior $\omega_{de}(z\rightarrow-1) > -1.$ For (4/3, 0.05), $\omega_{de0}$=-0.78 and 
$\omega_{de}(z\rightarrow-1)> -1.$ When $\beta$ takes negative values, the equation of state parameter have following values. For 
(1.2,-0.1), $\omega_{de0}$=-0.96 and $\omega_{de}(z\rightarrow-1)<-1.$ For (4/3, -0.05), $\omega_{de0}$=-0.88 and 
$\omega_{de}(z\rightarrow-1)<-1.$ For (1.01, -0.01), $\omega_{de0}$=-0.96 and $\omega_{de}(z\rightarrow-1)=-1.$
The WMAP-7 data predicts $\omega_{de0} \sim -0.93$, when joint analysis of the WMAP+BAO+H$_0$+SN data \cite{Komatsu1,Luis1} for constraining the present 
value of the equation of state parameter for the dark energy is made. 
From our analysis, the present values of $\omega_{de}$ for the parameter sets (1.2,-0.1) and (1.01,-0.01) are very close to the 
observationally deducted values. However the parameters (1.2,-0.1) have the behavior that in the far future, as $z\rightarrow-1$, the 
$\omega_{de}$ crosses the phantom divide -1, so the model posses phantom behavior in the future evolution. While for the parameter set
(1.01,-0.01), the $\omega_{de}$ approaches -1, which corresponds the $\Lambda$CDM model where the energy density is fully dominated
with the cosmological constant. So the parameter set $(\alpha,\beta)$=(1.01,-0.01) can be considered as preferable values.
The plots also shows that irrespective of the values of the parameters, $\omega_{de} \rightarrow 0$, at very large redshift. That is in the 
remote past the IMHRDE1 is behaving almost like pressureless cold dark matter for all values of the parameter set.

\begin{figure}[h]
\includegraphics{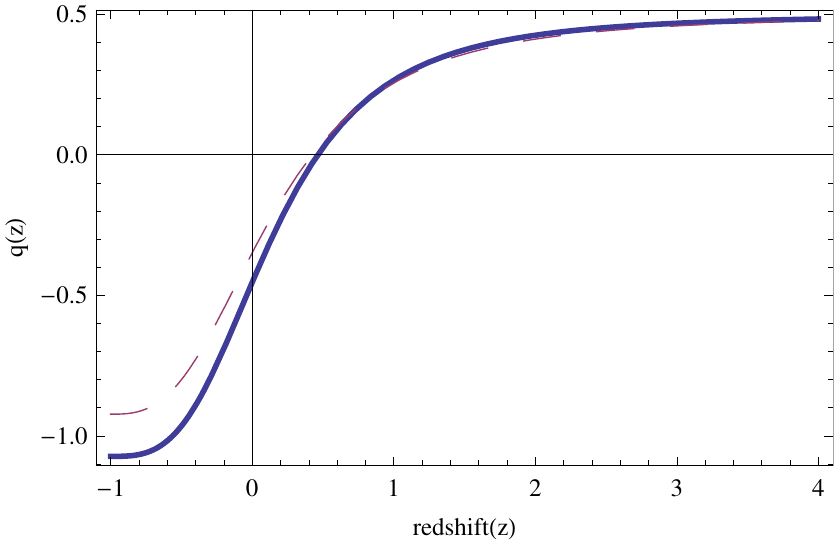}
\caption{Behavior of the deceleration parameter with redshift for parameters $(\alpha,\beta)$=(4/3,0.05) (dashed line), 
(4/3,-0.05) (continuous line), with
coupling parameter $b$=0.001.}
\label{fig:q1}
\end{figure}

Apart form the Hubble parameter $H$, deceleration parameter $q,$ is another geometrical parameter which describes the expansion 
history of the universe. The deceleration parameter can be expressed in terms of $h$ as,
\begin{equation} 
 q = - \frac{1}{2h^2} {dh^2 \over dx} - 1
\end{equation}
Accelerated expansion is indicated by the condition $q<0.$ From equation \ref{eqn:h2}, $q$-parameter can be expressed as,
\begin{equation} \label{eqn:q1}
q=-{3 \left(c_1 m_1 e^{\frac{3}{2} m_1 x} + c_2 m_2 e^{\frac{3}{2} m_2 x} \right) \over 
4 \left(c_1 e^{\frac{3}{2} m_1 x} + c_2  e^{\frac{3}{2} m_2 x} \right) } - 1
\end{equation}
In the non-interaction limit and by avoiding the contribution from dark matter, the $q$-parameter becomes $q=(3\beta -2)/2,$ 
which is in agreement with 
our earlier work \cite{tkm1} and it shows, as $\beta$ increases form zero, the $q-$parameter increases from -1. In figure \ref{fig:q1} 
we have plotted the evolution of $q-$parameter of the interacting MHRDE with redshift.

\begin{figure}[h]
 \includegraphics{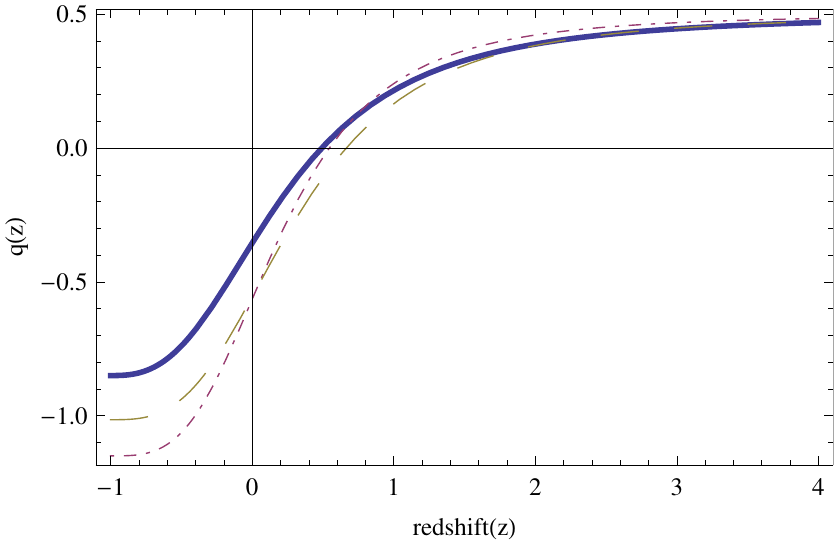}
\caption{Evolution of $q-$parameter with redshift for parameter sets $(\alpha, \beta)$=(1.2,0.1)(thick continuous) ,
(1.2,-0.1)(dot-dashed ) and (1.01, -0.01) (dashed line) with
coupling constant $b$=0.001.}
\label{fig:q2}
\end{figure}

 For evolution of the $q$ parameter for different sets $(\alpha,\beta)$ parameters are shown in figures \ref{fig:q1} and \ref{fig:q2}. 
Both the plots shows the universe enters the accelerated expansion in the recent past, the corresponding to the transition to the 
accelerating phase are $z_T$=0.47 for parameters (4/3,-0.05), 0.55 for (1.2,-0.1), 0.70 for (1.01,-0.01), 0.44 for (4/3,0.05),
0.50 for (1.2,0.1) and 0.52 for (1.15,0.15). 
The combined 
analysis of SNe+CMB data with $\Lambda$CDM model gives the range for the redshift at which universe enters the accelerated phase is
$z_T$= 0.45 - 0.73 \cite{Alam1}. The transition redshift given above for the IMHRDE1 is seen to be in the observational range for almost 
all parameter sets. However in the case of predicting the equation of state, we have found that the parameter set (1.01,-0.01) giving the 
best suitable prediction. The transition redshift for this parameter set in $z_T=0.70$ is agreeing with the upper limit 
region of the observationally constraint range.
Form the plots the present value of the deceleration parameter is -0.46 corresponds to (4/3,-0.05), -0.55 for (1.2,-0.1),
-0.56 corresponds to (1.01,-0.01), -0.34 corresponds to (4/3,0.05), -0.35 corresponds to (1.2,0.1) and -0.33 corresponds to (1.15,0.15).
The plot shows that, the universe enters the accelerating expansion in the recent past, with redshift in the range $z_T$= 0.52 - 0.57 and the 
present deceleration parameter is in the range $q$= 0.33 - 0.58. So it can be concluded that, compared to the $\Lambda$CDM model, 
the universe entering the accelerating expansion at a relatively later in the IMHRDE1 model. The observationally constraint value of the 
deceleration parameter form WMAP data is $q_0=-0.60.$ From this we can conclude that it is the parameter set (1.01,-0.01) giving the 
best value for $q_0$ as 0,56, which very close to the observational value. The parameter set (1.2,-0.1) is also giving
a competent value for the present deceleration parameter. But the IMHRDE1 tends towards phantom behavior in the future corresponds 
this parameter set, so we consider the set (1.01,-0.01) as the best parameter set.

\subsubsection{Statefinder analysis}

Various dark energy models predicts $H >0$, $q<0$ at the present time for the universe. Therefore Hubble parameter and deceleration 
parameter cannot discriminate dark energy models. In this light, sahni et. al. \cite{sahni1} and Alam et. al. \cite{Alam1}, introduced 
statefinder parameters $(r,s)$, by using third order time derivative of the scale factor. These effectively distinguishes various 
dark energy models by removing the degeneracy between $H$ and $q.$ The definitions of these parameters is given in equation 
(\ref{eqn:rs}). In terms of $h^2$ statefinder parameters can recast as 
\begin{equation}
 r = \frac{1}{2h^2}{d^2h^2 \over dx^2} + \frac{3}{2h^2}{dh^2 \over dx} +1
\end{equation}
and 
\begin{equation}
 s = - \left( { \frac{1}{2h^2} {d^2h^2 \over dx^2} + \frac{3}{2h^2}{dh^2 \over dx} \over \frac{3}{2h^2}{dh^2 \over dx} + \frac{9}{2} } \right) 
\end{equation}
On substituting $h^2$ from equation (\ref{eqn:h2}), the above equations become,
\begin{equation}
 r= 1 + {9 \left(c_1 m_1^2 e^{\frac{3}{2}m_1 x} + c_2 m_2^2 e^{\frac{3}{2} m_2 x} \right) + 2 \left(c_1 m_1 e^{\frac{3}{2}m_1 x} + c_2 m_2 
e^{\frac{3}{2} m_2 x} \right) \over 8 \left(c_1 e^{\frac{3}{2}m_1 x} + c_2  e^{\frac{3}{2} m_2 x} \right) }
\end{equation} 
and 
\begin{equation}
 s = - \left( { (c_1 m_1^2 e^{\frac{3}{2} m_1 x} + c_2 m_2^2 e^{\frac{3}{2} m_2 x} ) + 2 (c_1 m_1 e^{\frac{3}{2} m_1 x} + 
c_2 m_2 e^{\frac{3}{2} m_2 x} ) \over 2 (c_1 m_1 e^{\frac{3}{2} m_1 x} + c_2 m_2 e^{\frac{3}{2} m_2 x} ) + 4 
(c_1  e^{\frac{3}{2} m_1 x} + c_2  e^{\frac{3}{2} m_2 x} ) } \right)
\end{equation}
In the limiting case of non-interacting MHRDE ($b$=0), and also avoiding the contribution from dark matter ($\Omega_{mo} \sim 0 $),
 the above equations reduces to $r=1 + (9\beta(\beta -1))/2$ 
and $s = \beta$, which are in agreement with the earlier results for the non-interacting case \cite{tkm1}. So under these specific 
condition, at $\beta =0$, the MHRDE model corresponds to the $\Lambda$CDM (LCDM) model with $(r,s)$=(1,0) in the $r-s$ plane.

\begin{figure}[h]
\includegraphics[scale=0.60]{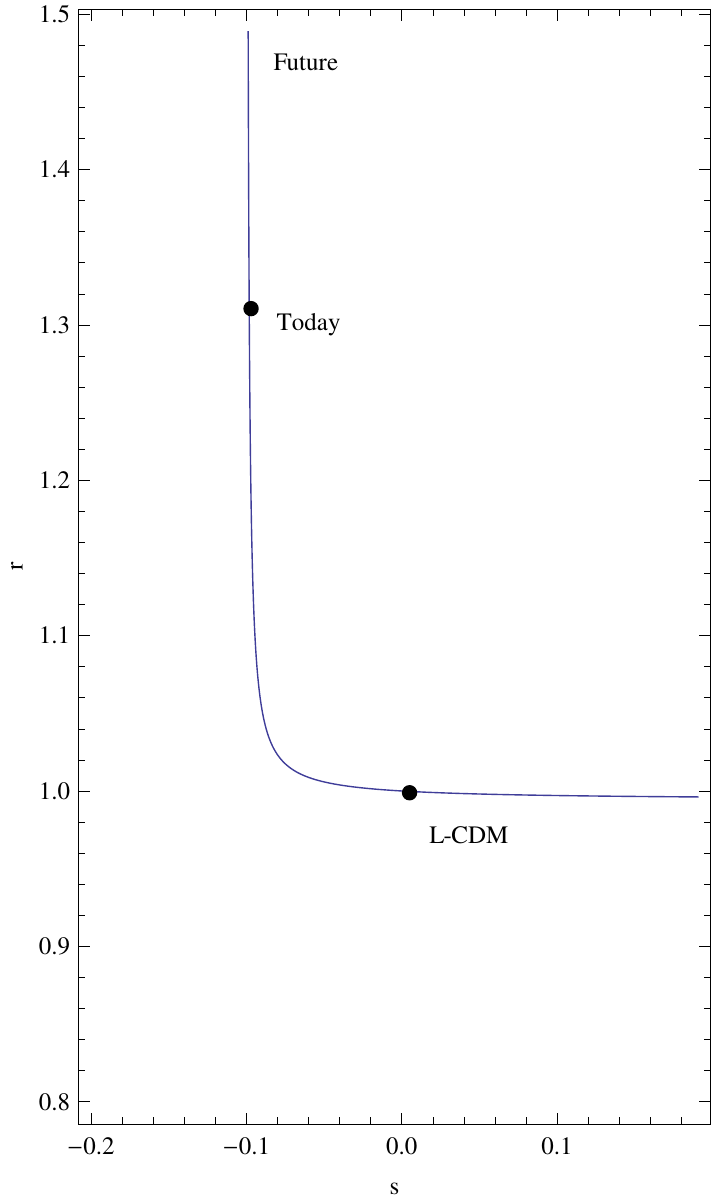}
 \includegraphics[scale=0.60]{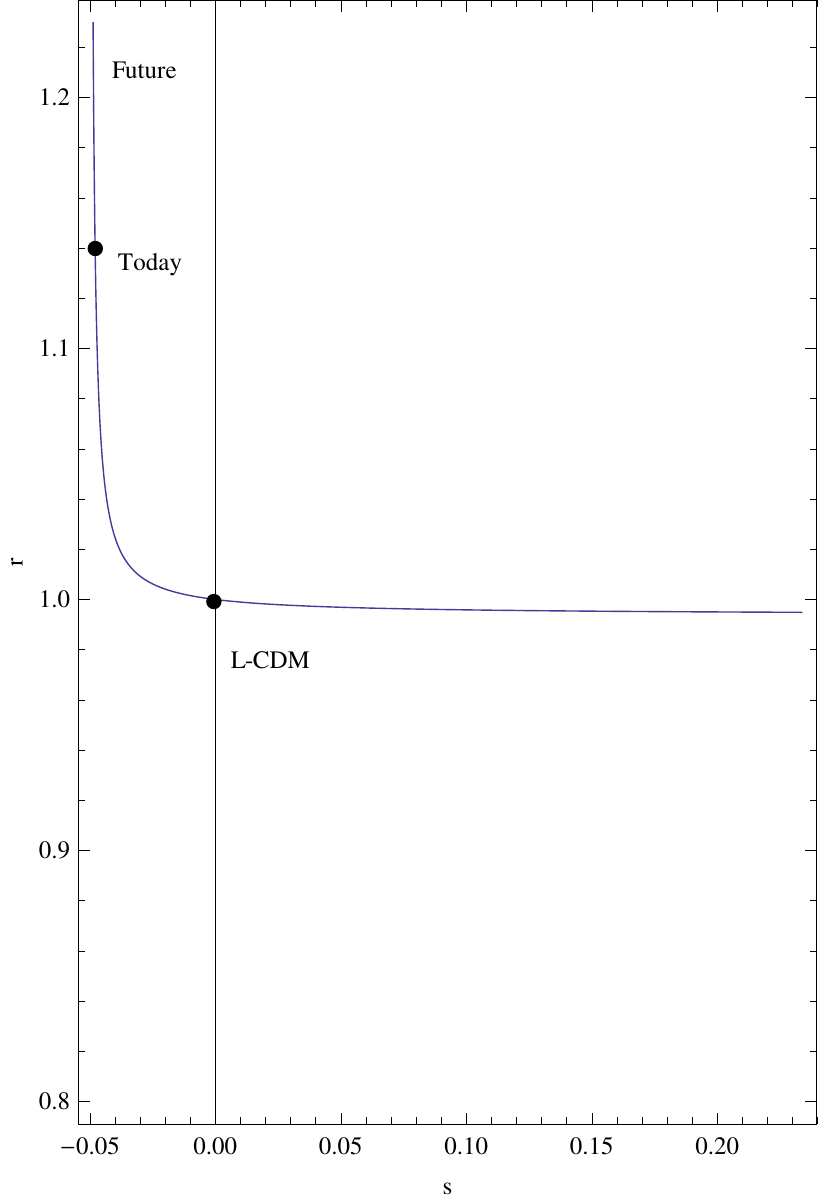}
\includegraphics[scale=0.60]{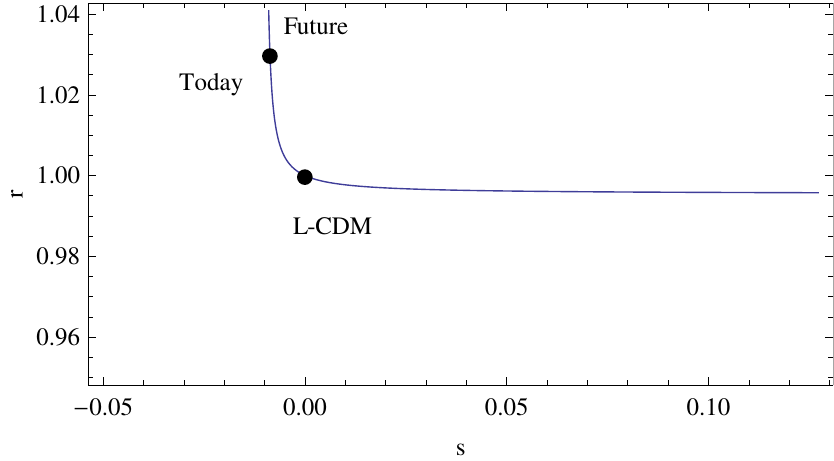}
\caption{$r-s$ evolutionary trajectories for the model for parameters, $(\alpha, \beta)$=(1.2,-0.1), (4/3, -0.05) and 
(1.01,-0.01)  with $b$=0.001}
\label{fig:rs1}
\end{figure}

The $r - s$ evolutionary trajectory of the interacting MHRDE model for the parameters $(\alpha, \beta)$ = (1.33,-0.05), (1.2, -0.1)
is shown in figure \ref{fig:rs1}
The evolution is starts from the right and evolves to the left in the $r-s$ plane. The plots shows that, the $r$ parameter stays almost
constant in the beginning stages of the expansion of the universe.
The $\Lambda$CDM phase, corresponds 
to $(r,s)$=(1,0), as denoted in the plots by the point LCDM.  The todays position of the universe is also noted in the $r-s$ plane. 
From figure \ref{fig:rs1} it is clear that the for negative values of $\beta$, the IMHRDE1 is evolves through the $\Lambda$CDM phase. 
After that the universe is evolving in such a way that the $r$ value increases very steeply. The present values $(r_0,s_0)$ for different 
parameters are, (1.14,-0.048) corresponds to $(\alpha,\beta)$=(4/3,-0.05), (1.31,-0.098) corresponds to parameters (1.2,-0.1) and 
(1.03,-0.008) corresponds to parameters (1.01,-0.01). The distance of the point $(r_0,s_0)$ from the $LCDM$ fixed point 
is the least for the parameters $(\alpha,\beta)$=(1.01,-0.01). From the figure \ref{fig:rs1}, it is seen that as $\beta$ increases 
the distance between the today's point and $\Lambda$CDM point is decreasing.
\begin{figure}[h]
 \includegraphics[scale=0.60]{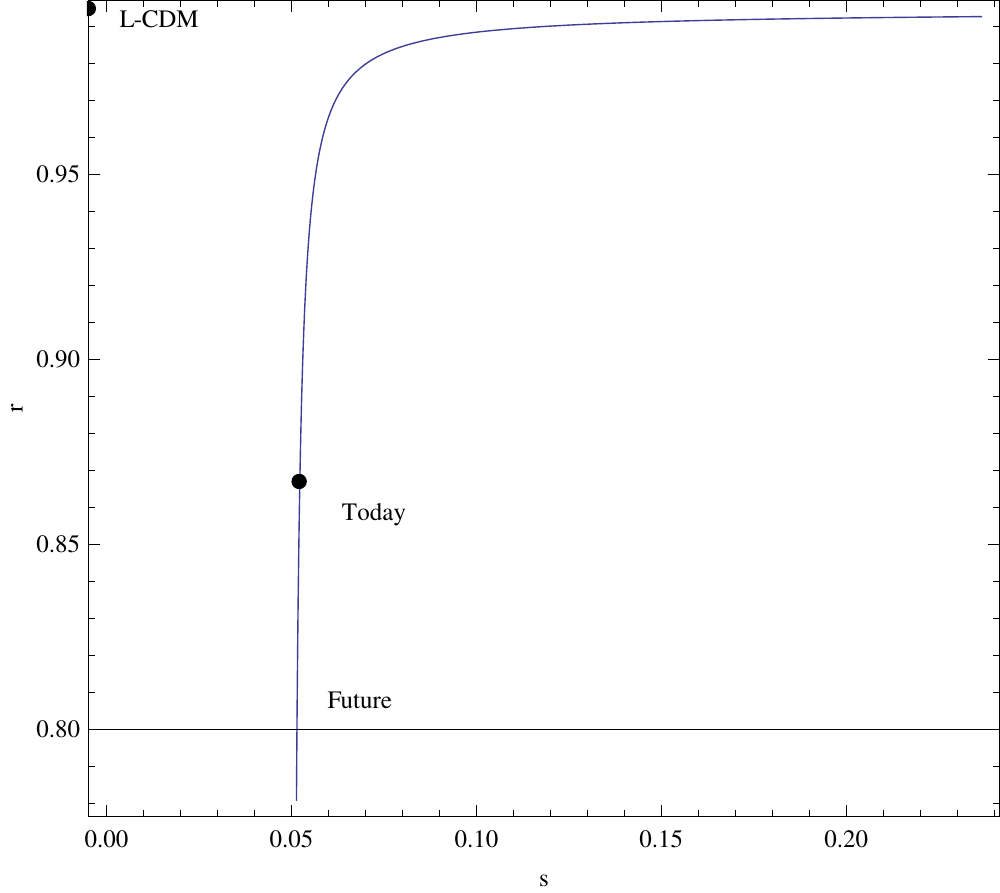}
\includegraphics[scale=0.60]{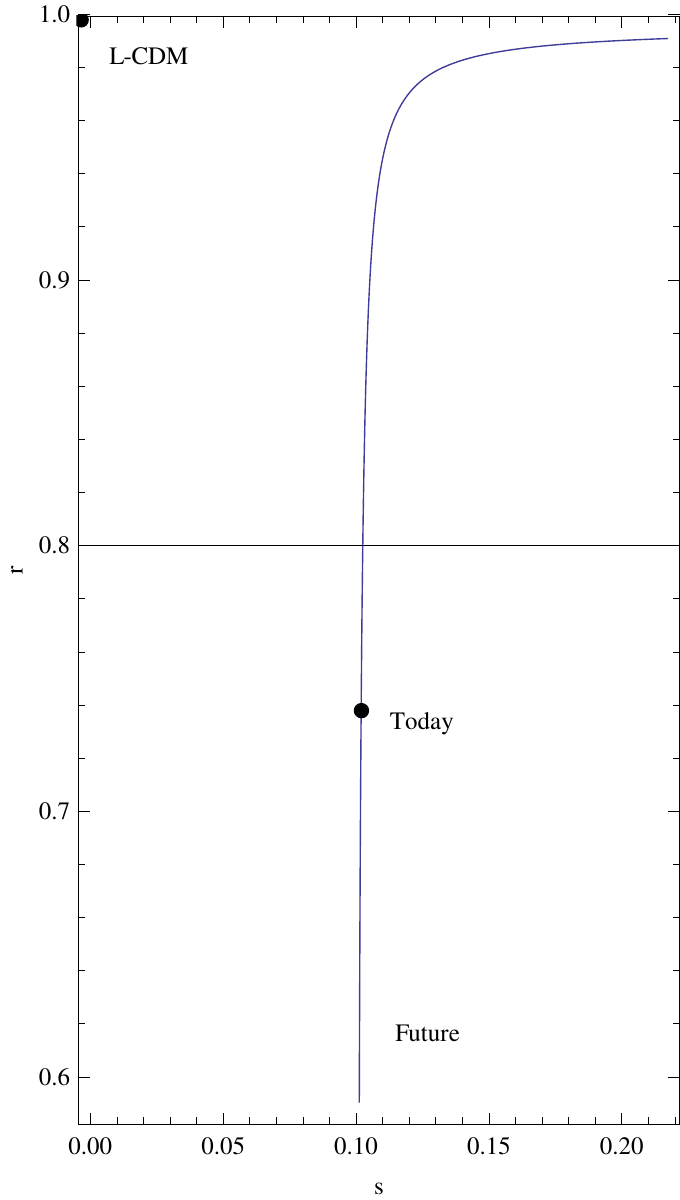}
\caption{$r-s$ evolutionary trajectories for parameters $(\alpha,\beta)$=(4/3,0.05), (1.2, 0.1), with $b$=0.001}
\label{fig:rs2}
\end{figure}
For 
positive $\beta$values the $r-s$ evolution is shown in figure \ref{fig:rs2}. Unlike in the case for negative $\beta$ values, here it is 
seen that the $\Lambda$CDM point is being a part of the $r-s$ evolution. Moreover in the later stages of evolution, the $r$ value 
decreases rather than increasing as in the case of positive $\beta$ values.  Another point to be noticed is that the distance between the
 $\Lambda$CDM point and today's 
point in the $r-s$ plane is increasing as $\beta$ increases.

The state finder diagnostic is clearly distinguishing the IMHRDE1 from other models. For quintessence model, the $r-s$ trajectory is 
lying in a region with $s>0$, $r<1$, for Chaplygin gas model the $r-s$ trajectory in the region with $r>1$, $s<0.$ For Holographic 
dark energy model with event horizon as the IR cut-off, the $r-s$ evolution starts it's evolution with $s=2/3$, $r=1$ and ends at the $\Lambda$CDM point in 
the $r-s$ plane. In the case of IMHRDE1, the distance between $\Lambda$CDM point and the today's point is the least in the $r-S$ plane 
for parameters $(\alpha,\beta)$=(1.01,-0.01), so this parameters can be favored over the other values. For further comparison with our 
results on IMHRDE1, one can see that in reference \cite{Setare1}, the authors have considered a new holographic dark energy model for 
which they have obtained $(r_0,s_0)$=(1.36,-0.59).

\begin{figure}[h]
 \includegraphics[scale=0.5]{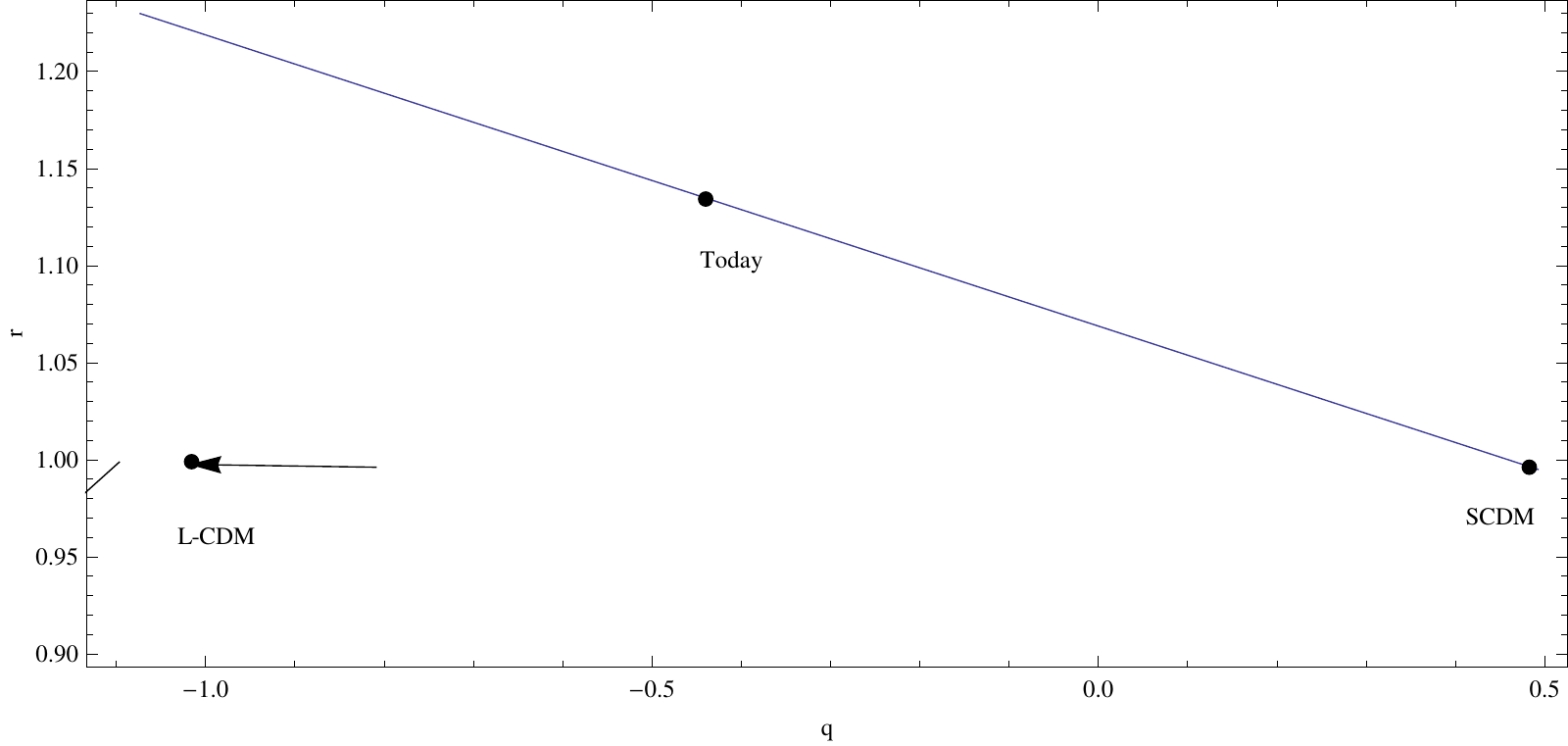}
\includegraphics[scale=0.5]{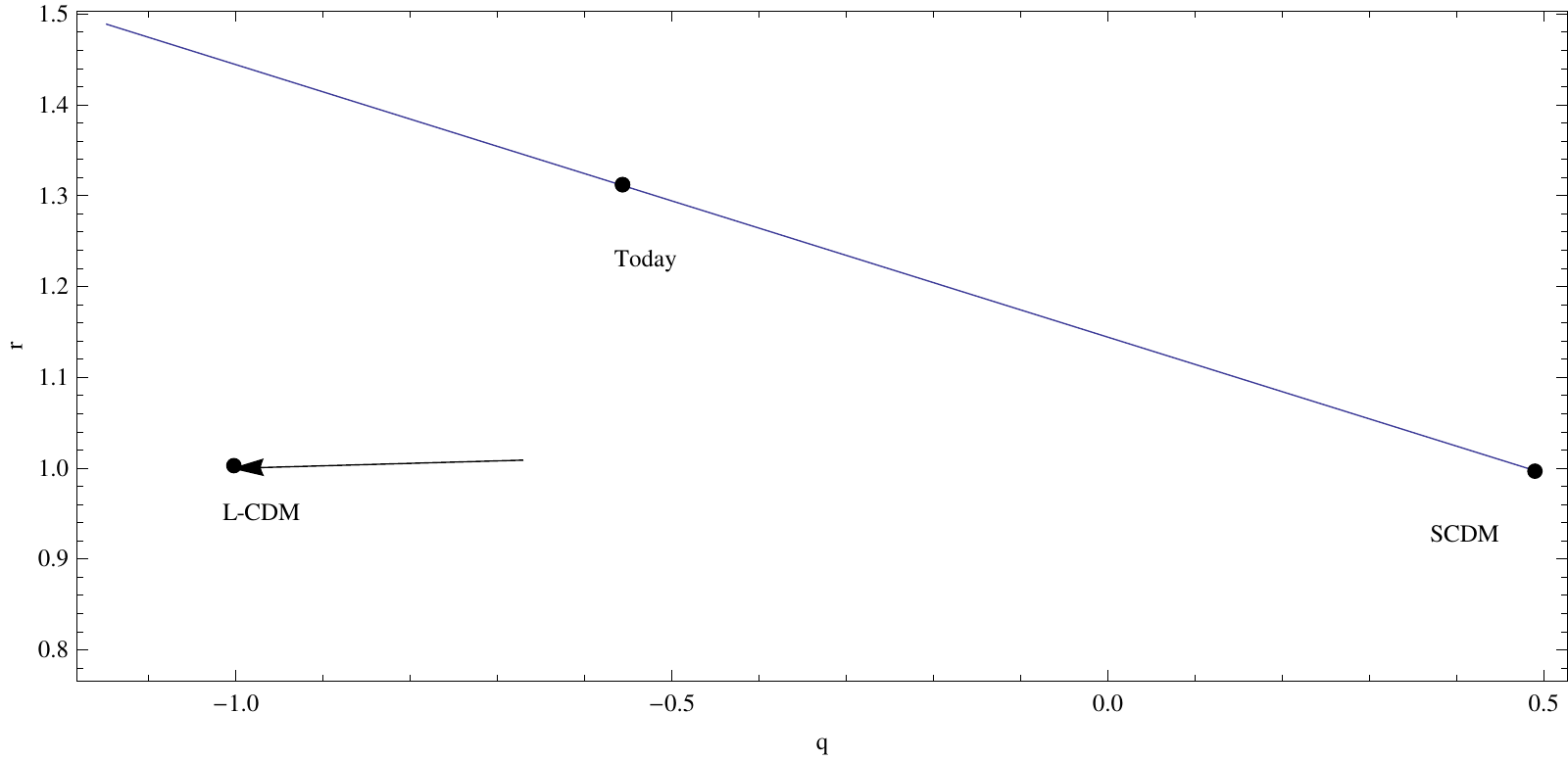}
\includegraphics[scale=0.5]{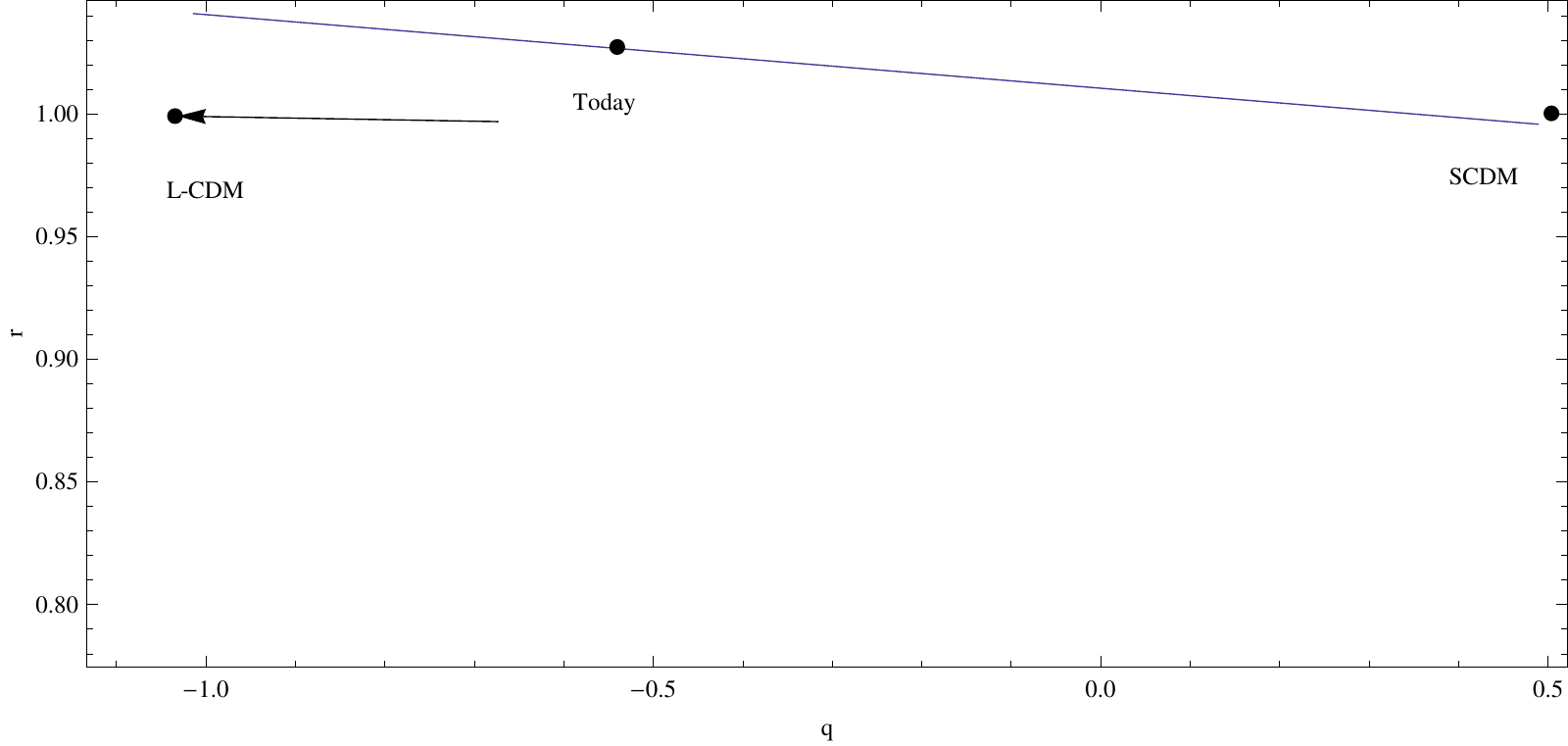}
\caption{$r-q$ plots for parameters $(\alpha, \beta)$=(1.33, -0.05), (1.2, -0.1), (1.01.-0.01) with interaction constant $b$=0.001}
\label{fig:rq1}
\end{figure}

\begin{figure}[h]
 \includegraphics[scale=0.5]{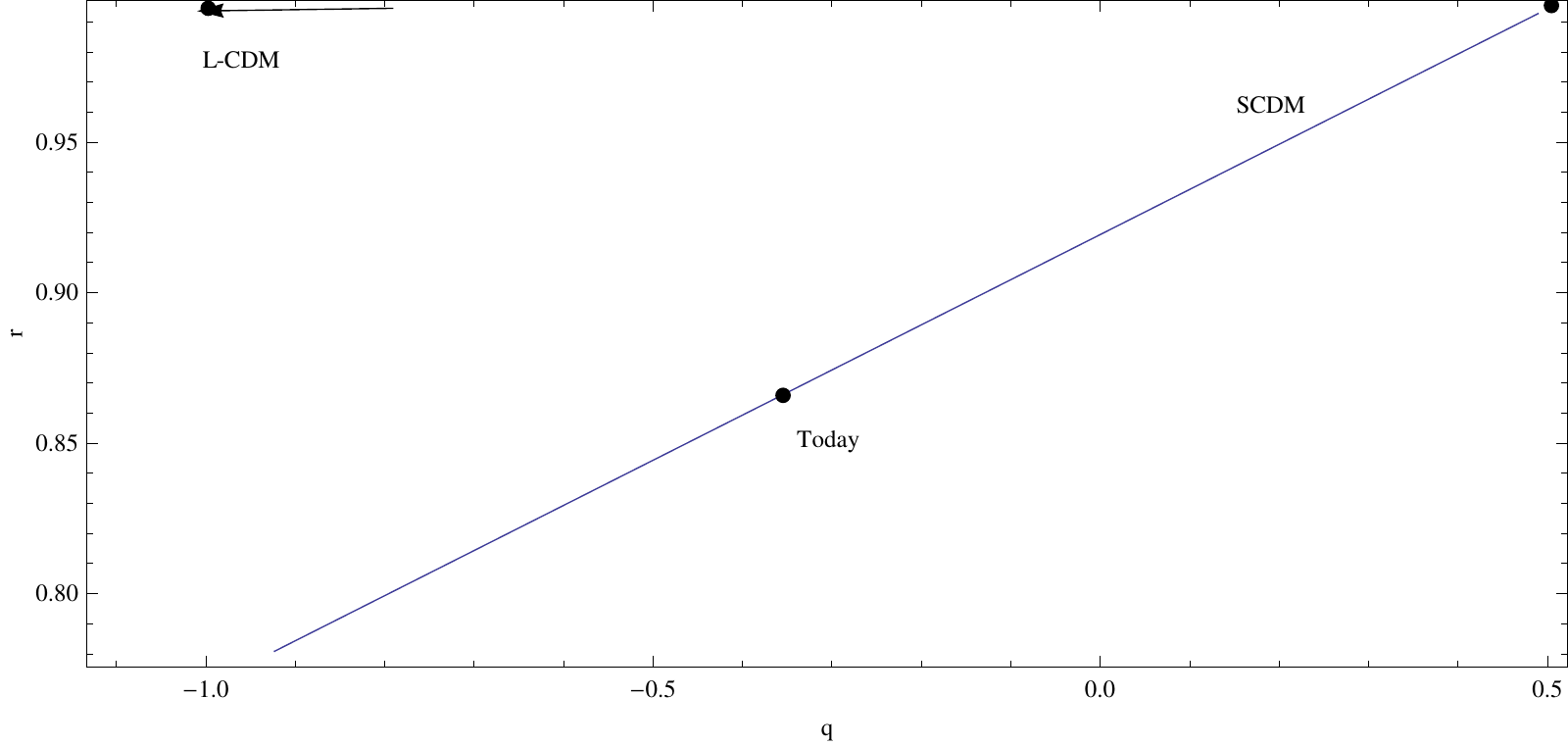}
\includegraphics[scale=0.5]{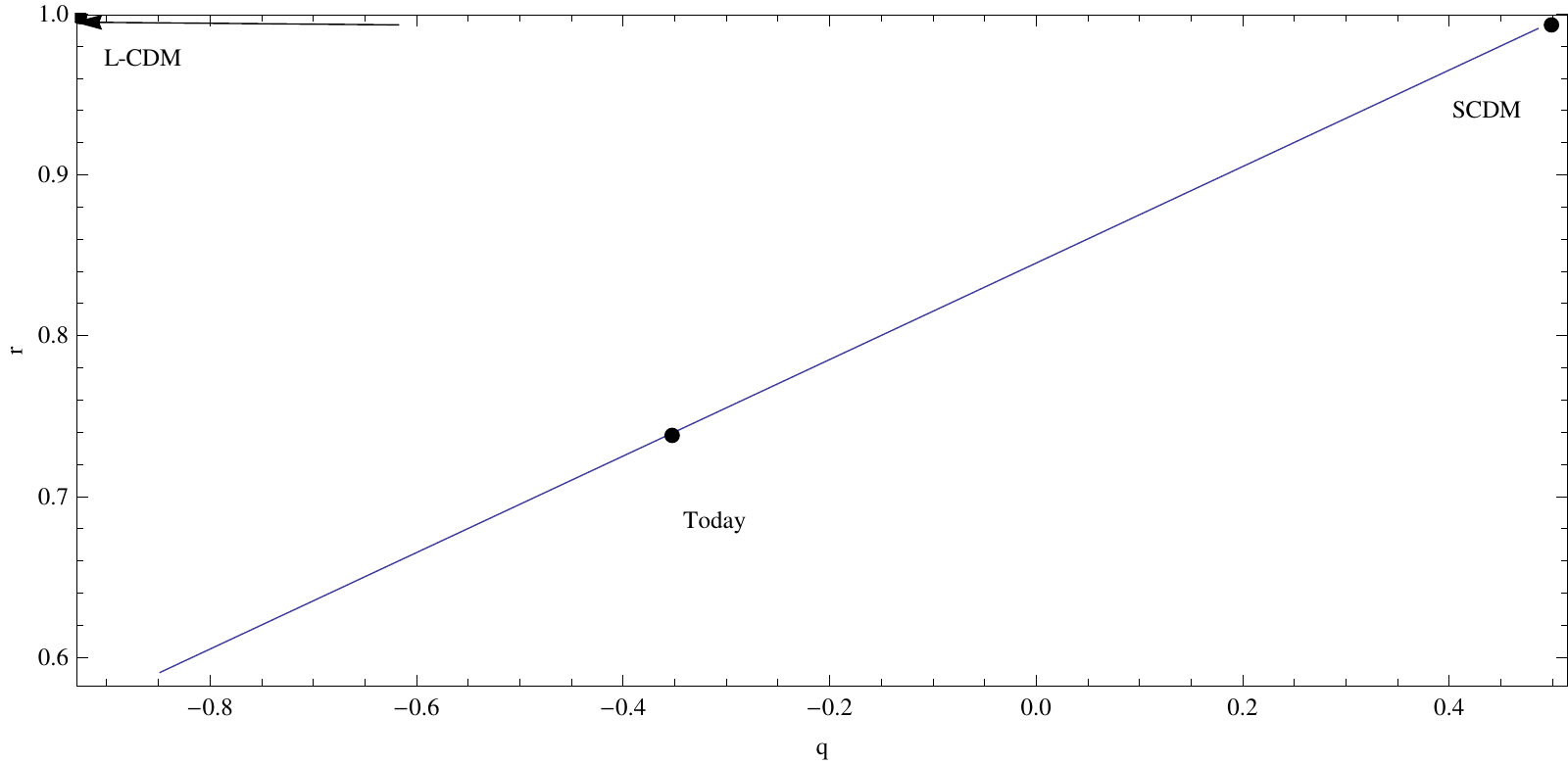}
\caption{$r-q$ plots for parameters $(\alpha, \beta)$=(1.33, 0.05), (1.2, 0.1), with interaction constant $b$=0.001}
\label{fig:rq2}
\end{figure}

We have studied the evolution of the IMHRDE1 model in the $r-q$ plane also. For  
interaction coupling constant  $b$=0.001, the plots are given figures \ref{fig:rq1} and \ref{fig:rq2}.

For negative values of $\beta$, the plots in figure \ref{fig:rq1} shows that  both $\Lambda$CDM model and 
IMHRDE1 model are commence evolving from the same point in the past corresponds to $r$=1, $q$=0.5, which corresponds to the matter 
dominated SCDM universe. For $\Lambda$CDM model the $r-q$ trajectory will end the evolution at $q$=-1,$r$=1, which corresponds to 
the de Sitter universe. The behavior of the IMHRDE1 is different from the above case. For the usual holographic 
dark energy model with event horizon as the IR cut-off, the starting and end point are similar to that of  the $\Lambda$CDM model
\cite{Zim1}. In the present model, even though it has the same starting point as the $\Lambda$CDM model, in the further evolution it is 
seemed to be different. For negative values of $\beta$ the $r$ value increase as $q$ decrease but for positive values of 
$\beta$ (fig. \ref{fig:rq2}),
the $r$ value decrease as $q$ decreases. The today's position of the universe in the $r-q$ plane is noted.
For negative values of $\beta$, the today's positions are $(r_0,q_0)$= (1.14, -0.45) corresponds to  $(\alpha, \beta)$=(4/3,-0.05), 
(1.31,-0.57) corresponds to (1.2,-0.1) and (1.03,-0.56) corresponds to (1.01,-0.01). For positive $\beta$ the today's position in the 
$r-q$ plane are (0.87,-0.35) corresponds to parameters (4/3,0.05), (0.74,-0.36) corresponds to (1.2,0.10) and (0.63,-0.33) corresponds to
parameters (1.15, 0.15). The parameters $(\alpha,\beta)$=(1.01,-0.01) can be considered as the best fit parameter, since it is giving the 
equation of state parameter almost same as deducted by the WMAP observations. For the same parameters, we have seen that $q_0$=-0.56, 
which is very near to the WMAP prediction -0.60 \cite{Komatsu1}. The $q_0$ value corresponds to the parameters (1.2,-0.1) is also compatible with
the corresponding WMAP values, moreover $\omega_{de0}$ corresponds to these parameters is -0.97. So the parameters (1,2,-0.1) is seems to
equally good as the parameters (1.01,-0.01), but the problem with these parameters is that the equation of state approaches value 
less than -1 as $z\rightarrow -1.$ As a result the equation of state parameter crosses the phantom divide and model leads to phantom 
behavior in the future evolution of the universe. In this light the parameters (1.01,-0.01) is finally preferred over other sets.

\subsection{Interacting model with $Q=3bH \rho_m$-IMHRDE2}
\label{sec:IMHRDE2}

\begin{figure}[h]
 \includegraphics[scale=0.75]{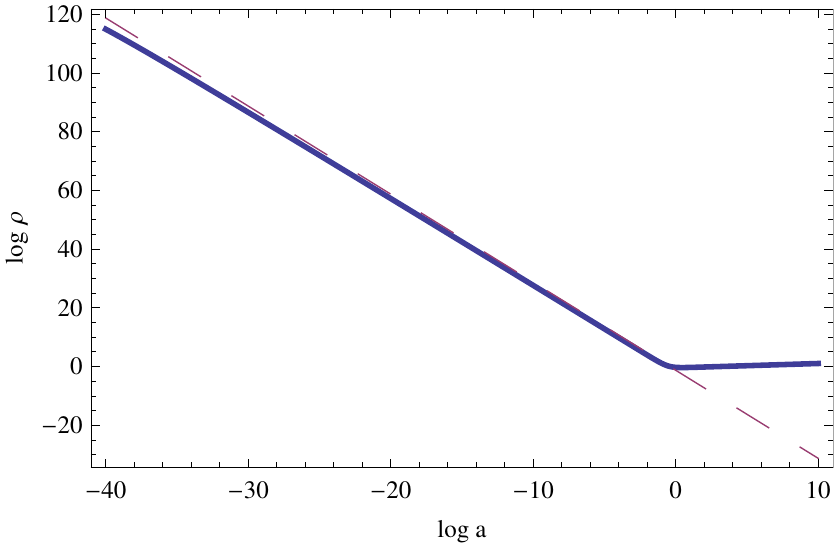}
\caption{Evolution of interacting MHRDE model with $Q=3bH\rho_m,$ for parameters, $(\alpha, \beta)$=(4/3, -0.05) and $b$=0.003. Continuous 
line for interacting MHRDE and dashed line is for dark matter. The plot shows that, dark energy is dominant in the recent past of the 
universe.}
\label{fig:coin2}
\end{figure}

In this section we are doing the same analysis of IMHRDE2 as in the previous section, but with interaction term given as, 
$Q=3bH\rho_m.$ From Friedmann equation the second order differential equation for $h^2$ can be obtained,
\begin{equation}
 {d^2h^2 \over dx^2} + 3 \left(\beta - b +1 \right) {dh^2 \over dx^2} + 9 \beta \left(1-b\right) h^2 =0
\end{equation}
A general solution for this can be writtes as,
\begin{equation}
 h^2 = k_1 e^{-3\beta x} + k_2 e^{3(b-1)x}
\end{equation}
where the constants $k_1$ and $k_2$ are evaluated using the initial conditions as, 
\begin{equation}
 c_1 = {\Omega_{de0} (\alpha - \beta) - \alpha - b +1 \over1- \beta - b  },  \, \, \, \, \, \, \, \, \, \, c_2 = 1-c_1
\end{equation}
 Again using Friedmann equation the dark energy density parameter can be obtained as,
\begin{equation}
 \Omega_{de} = k_1 e^{-3\beta x} + k_2 e^{3(b-1)x} - \Omega_{m0} e^{-3x}
\end{equation}
From this the dark energy pressure and equation of state parameter can be calculated as,
\begin{equation}
 p_{de}=- \left[ (1-\beta)k_1 e^{-3\beta x} + b k_2 e^{-3(1-b)x} \right]
\end{equation}
and
\begin{equation}
 \omega_{de}= -1 + \left[ { k_1 \beta e^{-3\beta x} + k_2 (1-b) e^{-3(1-b)x} - \Omega_{m0} e^{-3x} \over 
k_1 e^{-3\beta x} + k_2 e^{-3(1-b)x} - \Omega_{m0} e^{-3x} } \right]
\end{equation}
In the non-interaction case, with $b=0$ and $\Omega_{de0}=1,$ the coefficients become $k-1=1$ and $k_2=0$, then the the equation of 
state reduces to the standard form, $\omega_{de}=-1+\beta,$ confirming the 
earlier observations \cite{tkm1}. The value of the interaction coupling constant, $b$ is to be chosen in such a way that, the model 
must be viable with respect to coincidence between non-relativistic dark matter and dark energy. We found that the $b$ parameter can 
be around $b=0.003$ in the present case, at which the model 
explaining the coincidence problem very well as shown in figure \ref{fig:coin2}. For values $b>0.003$, the IMHRDE2 fails to be 
compatible with the co-evolution of dark energy and dark matter. So we take, $b$=0.003, for our further analysis in
this section.

\begin{figure}[h]
 \includegraphics[scale=0.75]{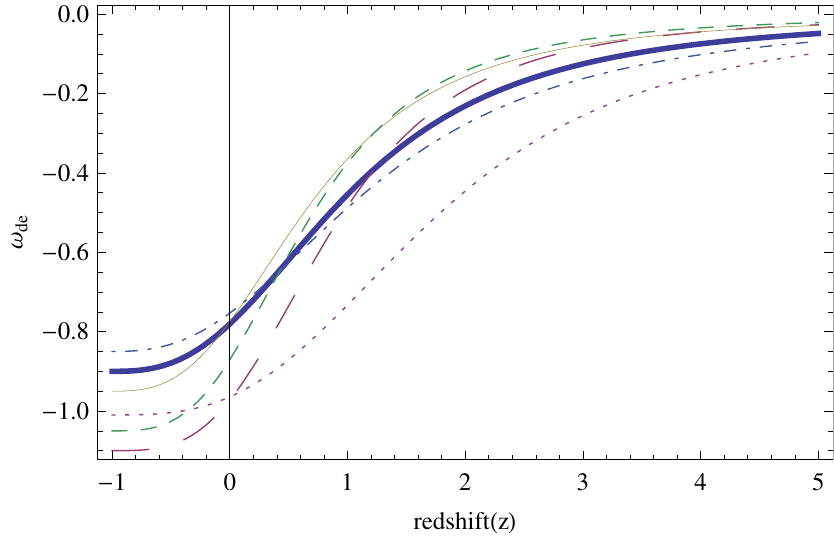}
\caption{Evolution of the equation of state for different parameters $(\alpha, \beta)$ with $b$=0.001. small dashed line is 
for $(\alpha,\beta)$=(1.15,0.15),long dashed line for (1.2, -0.1), dot-dashed line is for (4/3, -0.05), 
thin continuous line is for (4/3, 0.05) and thick continuous line is for (1.2, 0.1), doted line if for (1.01,-0.01)}
\label{fig:EOS21}
\end{figure}

The evolution of the equation of state parameter with redshift is shown in figure \ref{fig:EOS21}. The figure shows that, the equation of 
state parameter, is approaching zero, at very large positive values of redshift. Hence in the remote past, the interacting MHRDE is behaves 
like cold dark matter.
 As the universe evolves, the equation of state parameter become more and more negative, and approaches stabilization 
as $z \rightarrow -1.$ For parameters $(\alpha, \beta)$ = (4/3, -0.05), (1.2, -0.1), the $\omega_{de}$ approaches values below -1, at 
which it behaves as phantom dark energy. But for parameters $(\alpha, \beta)$=(4/3, 0.05), (1.2, 0.1), the equation of state saturate
at values above -1. For the best fit $(1.01,-0.01)$ the equation of state approaches -1 as $z\rightarrow -1.$ 
The present value of the equation of state parameter, is -0.88 for the parameters (4/3,-0.05), -0.98 for (1.2,-0.1), -0.97 
for (1.01,-0.01), -0.78 for (4/3,0.05), -0.78 for (1.2,0.1) and -0.76 for (1.15,0.15). It is seen that in contrast to the WMAP value, 
$\omega_{de0}$ corresponds to (1.2,-0.1) and (1.01,-0.01) are best values. As we mentioned earlier we will consider (1.01,-0.01) as 
the best parameters, which give the present equation of state as -0.97.

\begin{figure}[h]
 \includegraphics[scale=0.75]{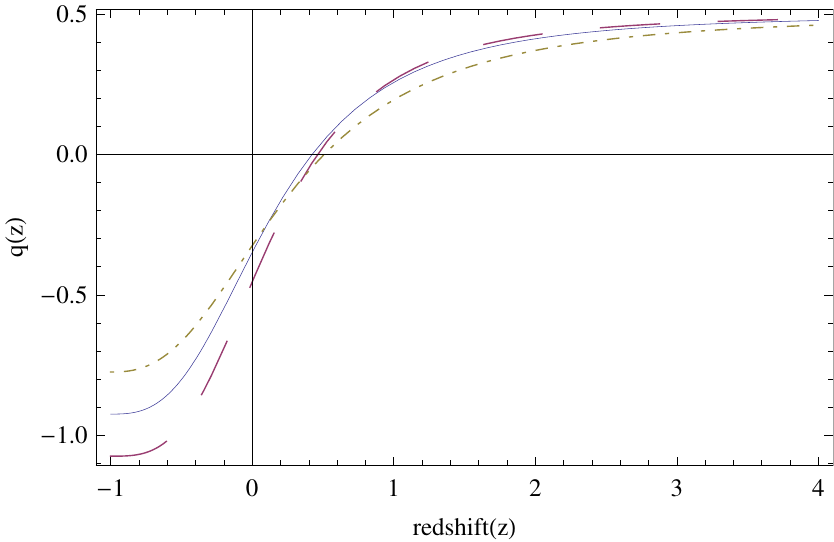}
\includegraphics[scale=0.75]{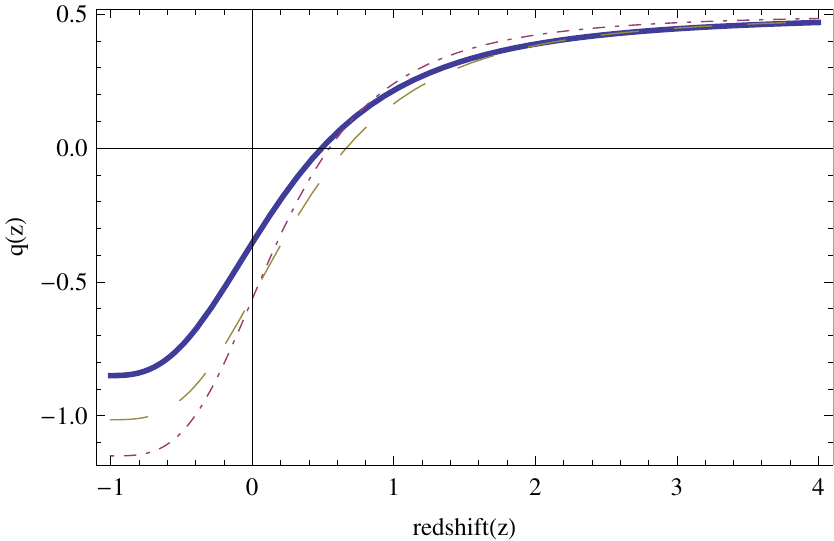}
\caption{Evolution of the deceleration parameter $q$ with redshift for parameters $(\alpha,\beta)$=(1.15,0.15)(dash-dot line), 
(1.33, -0.05)(dashed line), 
(1.33, 0.05)(continuous line), in the 
left and (1,2,-0.1)(dash-dot line), (1.2, 0.1)(continuous line), (1.01,-0.01)(dashed line) in the right, both for $b$=0.001.}
\label{fig:q212}
\end{figure}

The deceleration parameter $q$ in this case if found to be of the form
\begin{equation}
 q= -1 + \frac{3}{2} \left( { c_1 \beta e^{-3\beta x} - c_2 (b-1) e^{-3(1-b)x} \over c_1 e^{-3 \beta x} + c_2 e^{-3(1-b) x} } 
\right)
\end{equation}
In the non-interacting limit with negligible contribution form dark matter sector, the above equation reduces to $q=(3\beta -2)/2.$
We have studied the the evolution of the q parameter with redshift, and is shown in the figure \ref{fig:q212}.
The plot reveals that, at large redshift the $q-$parameter saturates at around  0.5. As the universe evolves, $q$ parameter starts 
decreasing and entering the negative value region corresponds to accelerating universe. The transition to the accelerating phase is 
occurred at $z_T$=0.44 corresponds to $(\alpha,\beta)$=(4/3,0.05), 0.50 corresponds to (1.2,0.1). For comparison the range deduced using 
observational data is $z_T$=0.45 - 0.73 \cite{Alam1}. This shows that, the present model IMHRDE2 is agreeing with the observational result 
for the parameters (1.2,-0.1) and (1.01,-0.01) with interaction coupling constant $b$=0.003. For the best fit parameters (1.01,-0.01) the 
the transition occurred at around $z_T$=0.68. It is also seen from the plots that, the present value corresponds to different sets 
parameters are -0.34 for (4/3,0.05), -0.36 for (1.2,0.1), -0.33 for (1.15,0.15), -0.46 for (4/3,-0.05), -0.57 for (1.2,-0.1) and 
-0.57 for (1.01,-0.01). For the best fit parameters (1.01,-0.01) the $q_0$=-0.57 is very much near to the observational 
prediction ) -0.60.

\subsubsection{Statefinder analysis}

In the following we will analyses the evolution of IMHRDE2 in the $r-s$ plane. For the present case these parameters, are given as 
\begin{equation}
 r= 1+ { 9c_1\beta^2 e^{-3\beta x}+ 9 c_2(1-b)^2 e^{-3(1-b)x}-9c_1 \beta e^{-3\beta x} -9c_2(1-b) e^{-3(1-b)x} \over 
2 (c_1 e^{-3\beta x} + c_2 e^{-3(1-b)x} ) }
\end{equation}
and
\begin{equation}
 s=- \left[ { 9c_1\beta^2 e^{-3\beta x} + 9c_2(1-b)^2 e^{-3(1-b)x} - 9c_1\beta e^{-3\beta x} - 9 c_2(1-b)e^{-3(1-b)x} \over
 -9c_1\beta e^{-3\beta x} - 9c_2(1-b) e^{-3(1-b)x} + 9c_1 e^{-3\beta x} + 9 c_2(1-b)e^{-3(1-b)x} } \right]
\end{equation}
In the non-interacting limit, the above parameters become, $r=1+9\beta (\beta - 1)/2$ and $s=\beta,$ which confirms the earlier results.
In this limit $r$ decreases as $\beta$(provided it is positive) increases, while for negative $\beta$ the $r$ parameter decreases as 
$\beta$ increases. In this model with coupling constant $b$=0.003, the evolution in the $r-s$ plane is as shown in the 
following figures \ref{fig:rs211} and \ref{fig:rs212}
\begin{figure}[h]
 \includegraphics[scale=0.5]{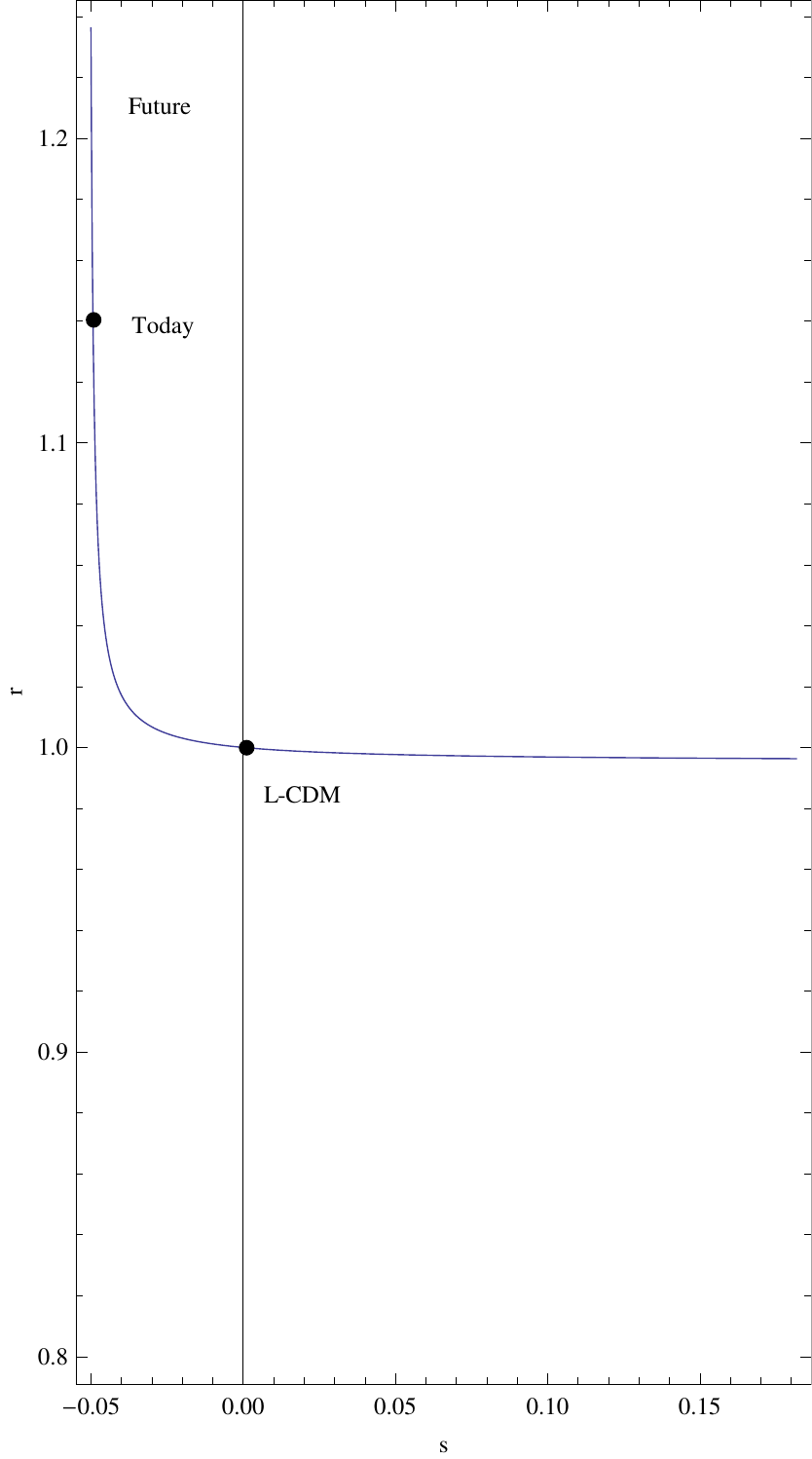}
\includegraphics[scale=0.5]{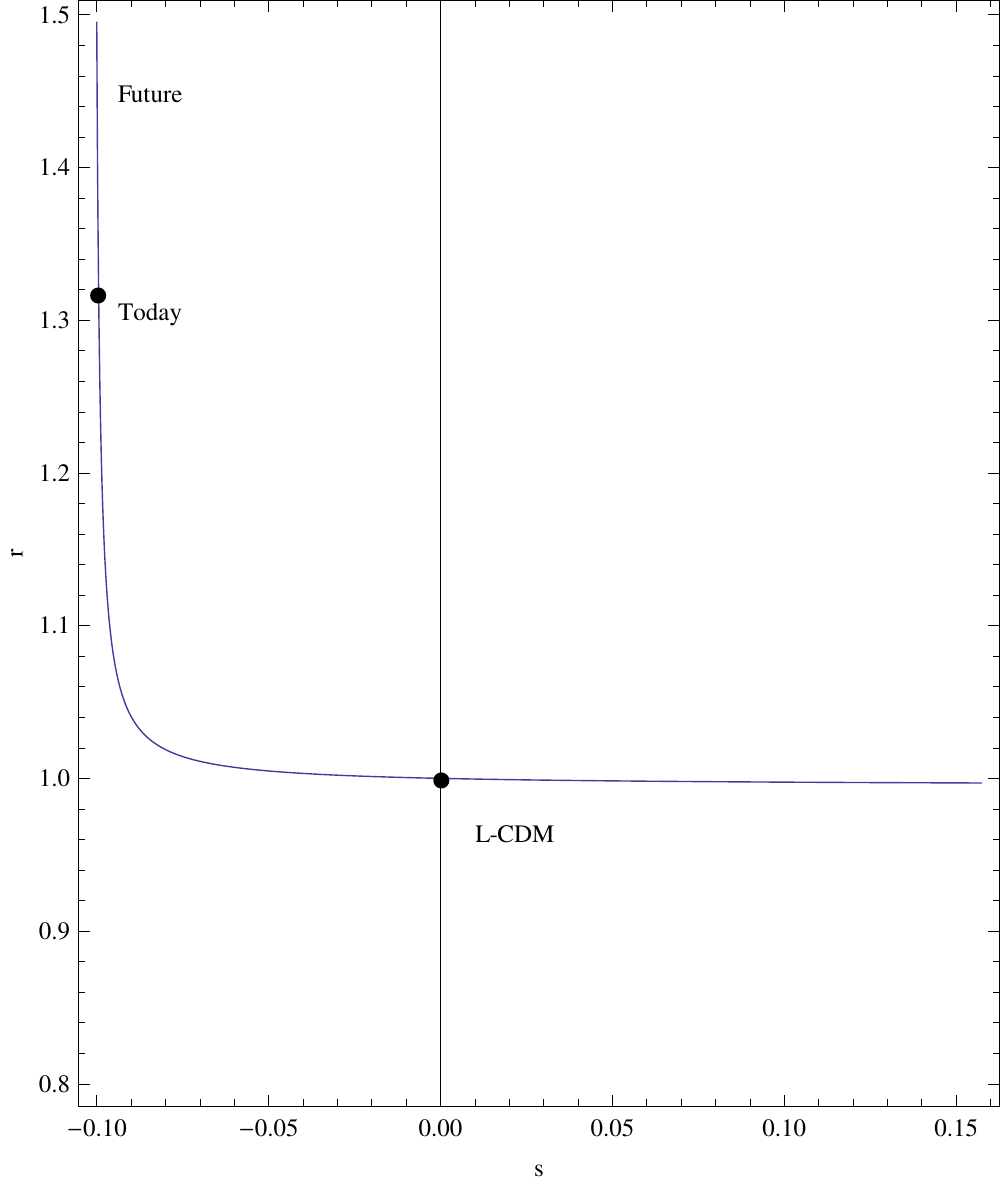}
\includegraphics[scale=0.5]{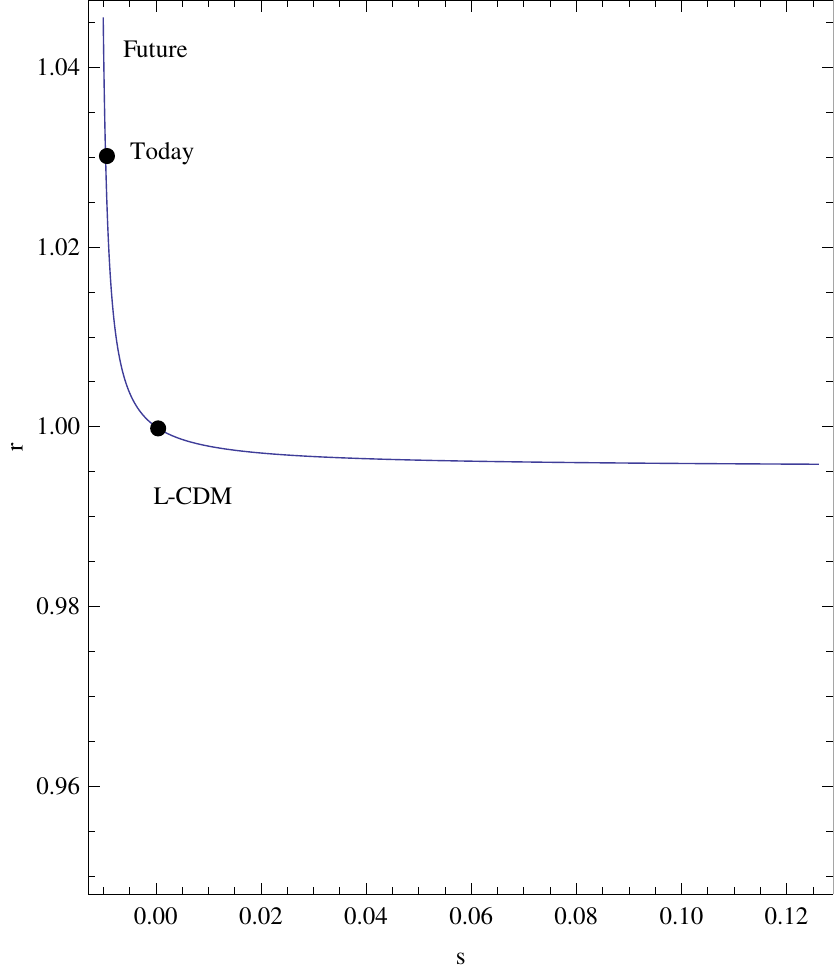}
\caption{Evolution of the model in the $r-s$ plane $(\alpha,\beta)$=(1.33,-0.05),(1.2, -0.1),(1.01,-0.01) with interaction 
coupling constant $b$=0.003}
\label{fig:rs211}
\end{figure}

\begin{figure}[h]
\includegraphics[scale=0.5]{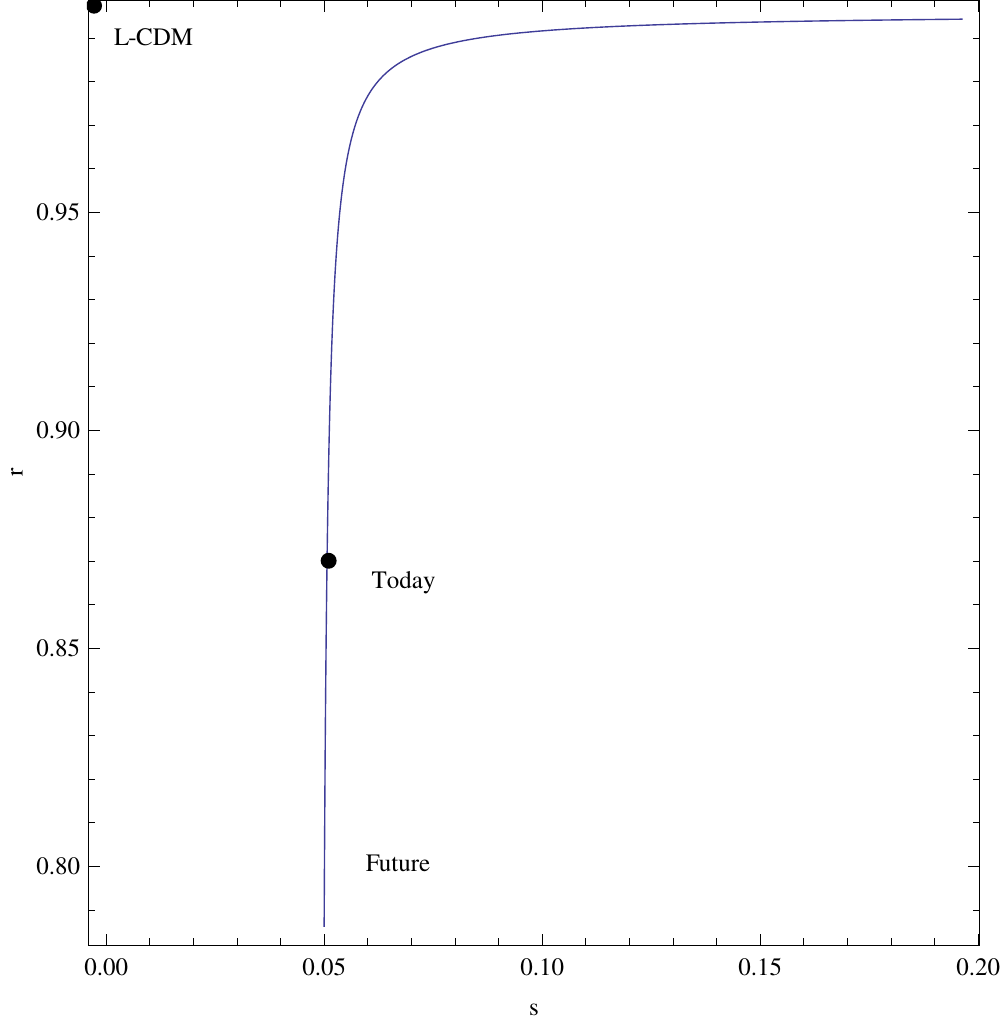}
\includegraphics[scale=0.5]{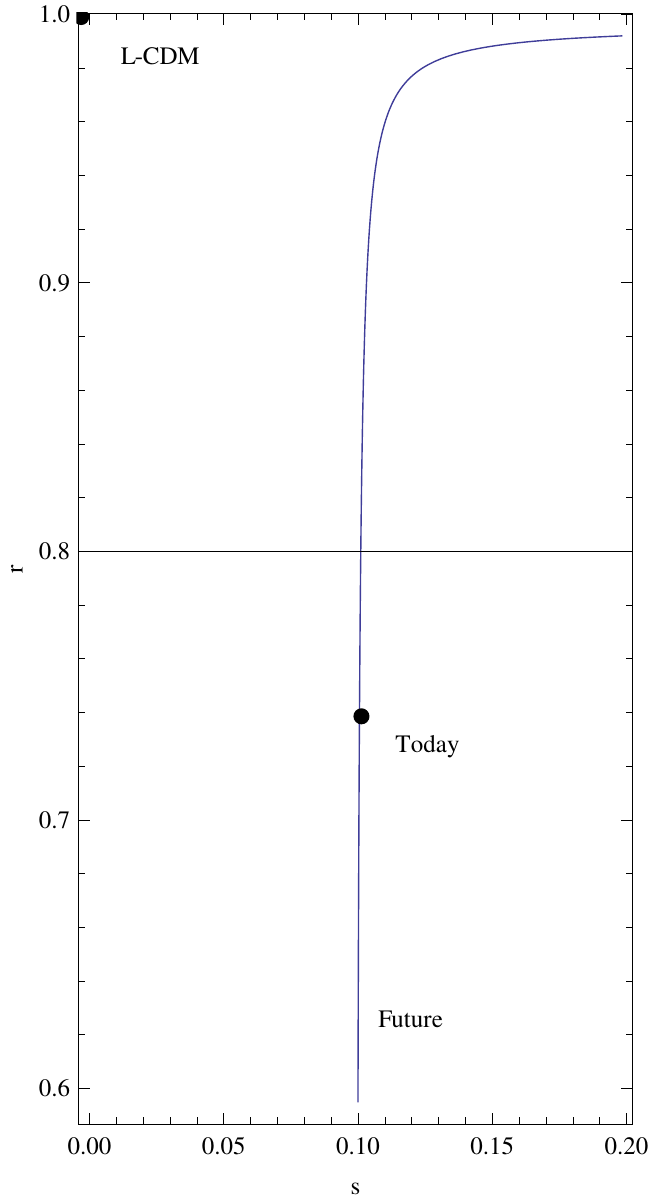}
\includegraphics[scale=0.5]{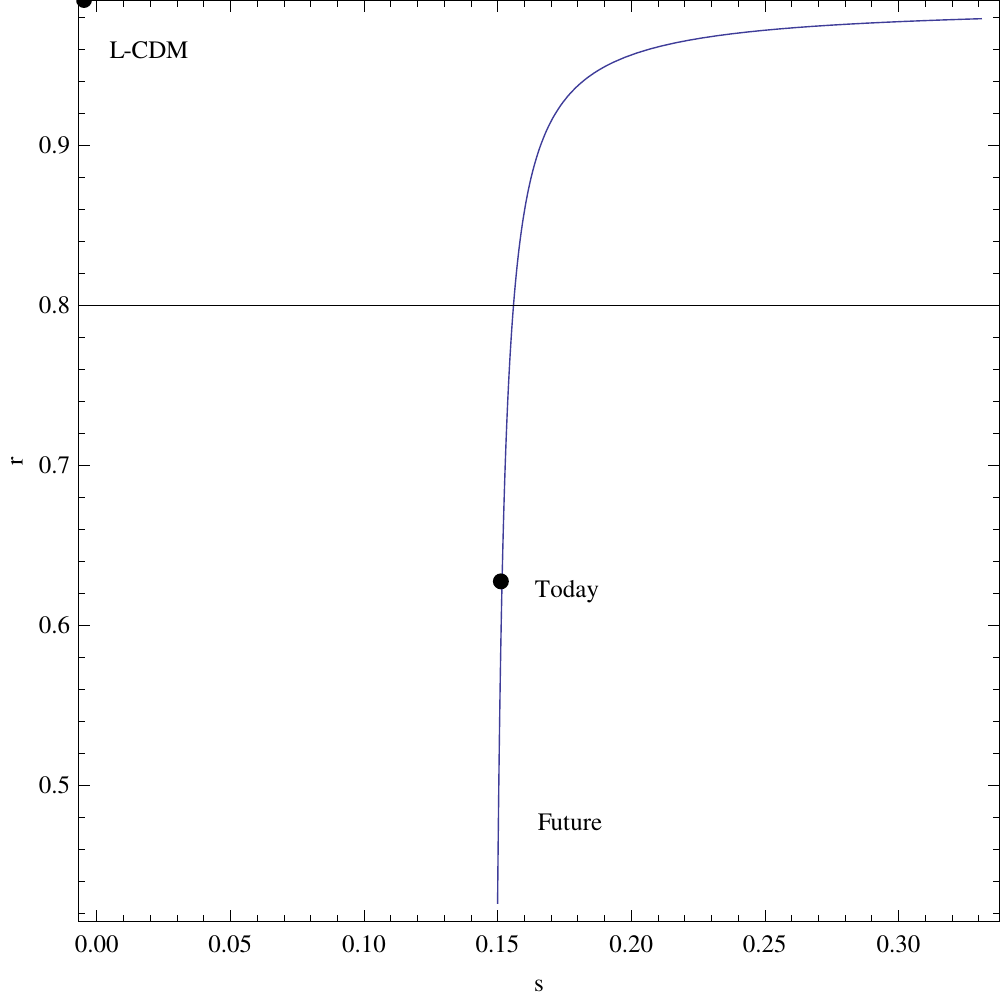}
\caption{Evolutionary trajectories of $r-s$ plane for parameters $(\alpha,\beta)$=(1.33,0.05),(1.2,0.1), (1.15,0.15) with $b$=0.003}
\label{fig:rs212}
\end{figure}
and it is seen that, irrespective of the values of parameters, in  IMHRDE2, the universe begin with
$r$=1 in the past. For negative values of $\beta$, the model is evolving through the $\Lambda$CDM model in the past.
 As $\beta$ increases the gap between $\Lambda$CDM point and the present IMHRDE2 phase is 
decreasing in the $r-s$ plane. For negative $\beta$, the present values of the parameters are $(r_0, s_0)$=(1.14, -0.048) corresponds to
(4/3,-0.05), (1.31,-0.099) corresponds to (1.2,-0.1) and (1.03,-0.0096) corresponds to (1.01,-0.01). 
For positive $\beta$ values the distance between the $\Lambda$CDM point and the today's point is increasing as $\beta$ increases.  
The present values of the parameters in these cases are $(r_0,s_0)$=(0.87, 0.05) corresponds to (4/3,0.05), (0.74, 0.10) corresponds to 
(1.2,0.1) and (0.62,0.15) corresponds to (1.15,0.15). For the best fit parameters $(1.01,-0.01)$ the present universe is corresponds to
$(r_0,s_0)$=(1.03,0.0096). So the $r-s$ parameter for the present universe clearly distinguishing IMHRDE2 from the $\Lambda$CDM model with
$(r_0,s_0)$=(1,0) and also from the new holographic dark energy model with $(r_0,s_0)$=(1.36,-0.102) \cite{Setare1}.

\begin{figure}[h]
\includegraphics[scale=0.65]{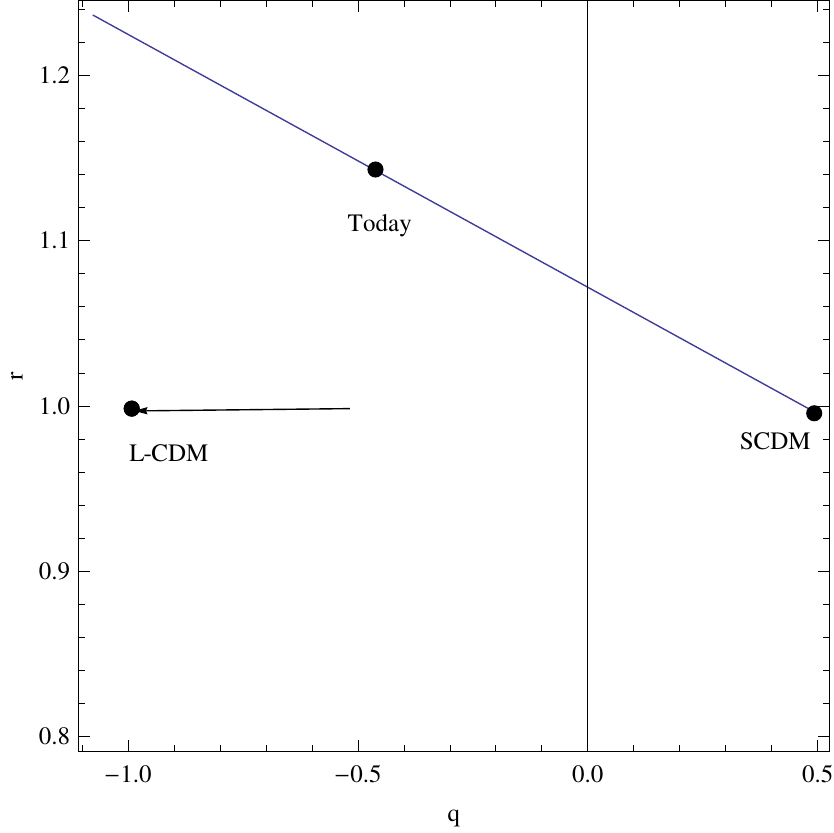}
\includegraphics[scale=0.65]{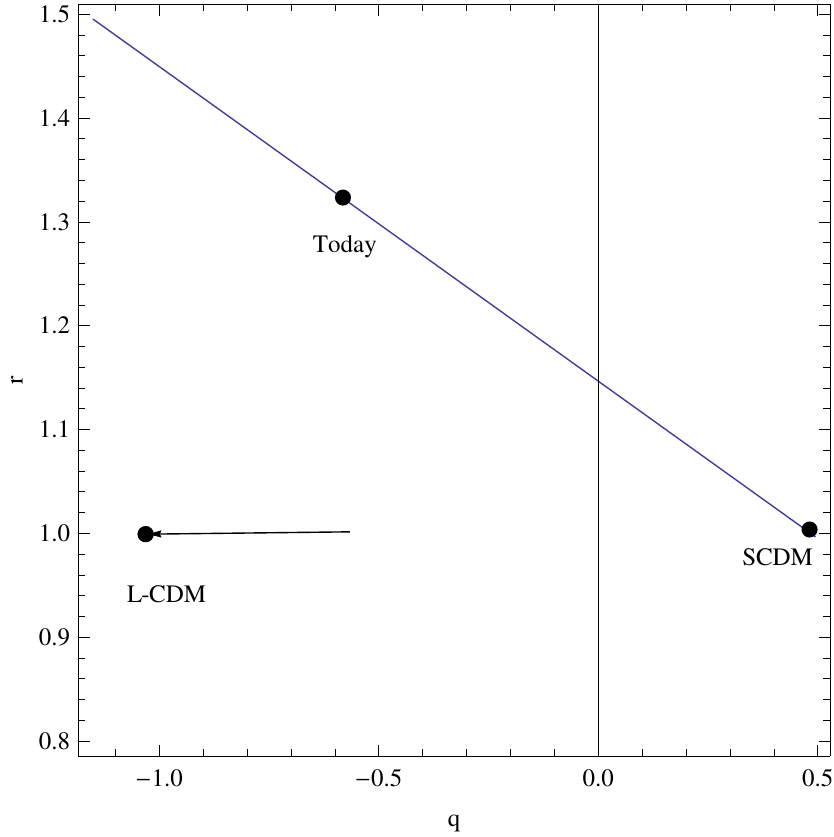}
\includegraphics[scale=0.65]{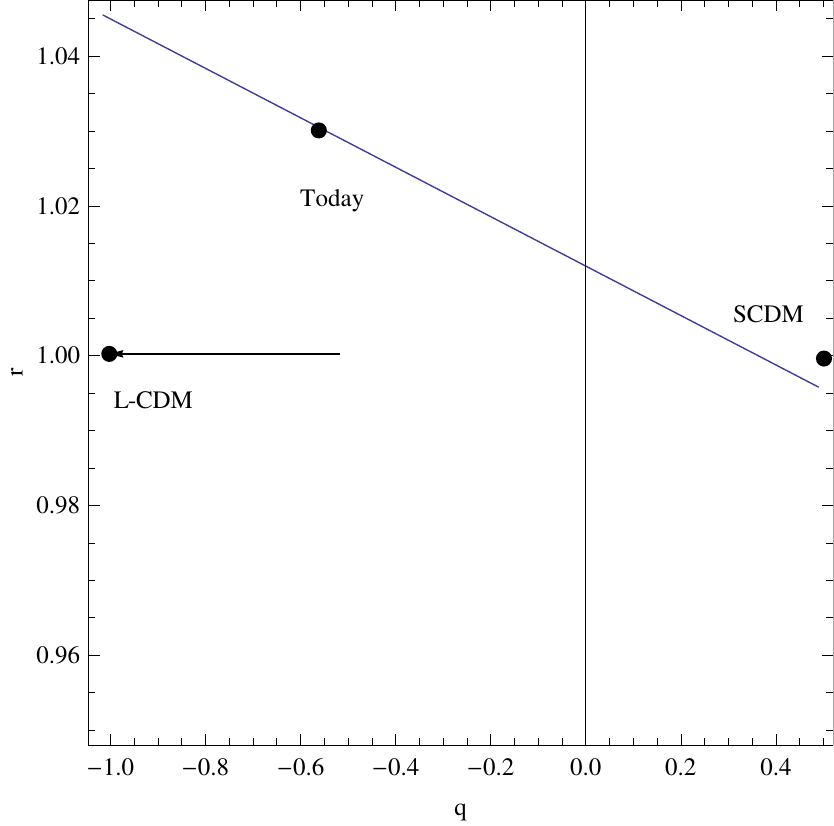}
\caption{Evolution of the interactive MHRDE model in the $r-q$ plane for parameters $(\alpha, \beta)$=(4/3, -0.05), (1.2, -0.1),
(1.01,-0.01) with $b$=0.001}
\label{fig:rq3}
\end{figure}

\begin{figure}[h]
\includegraphics[scale=0.65]{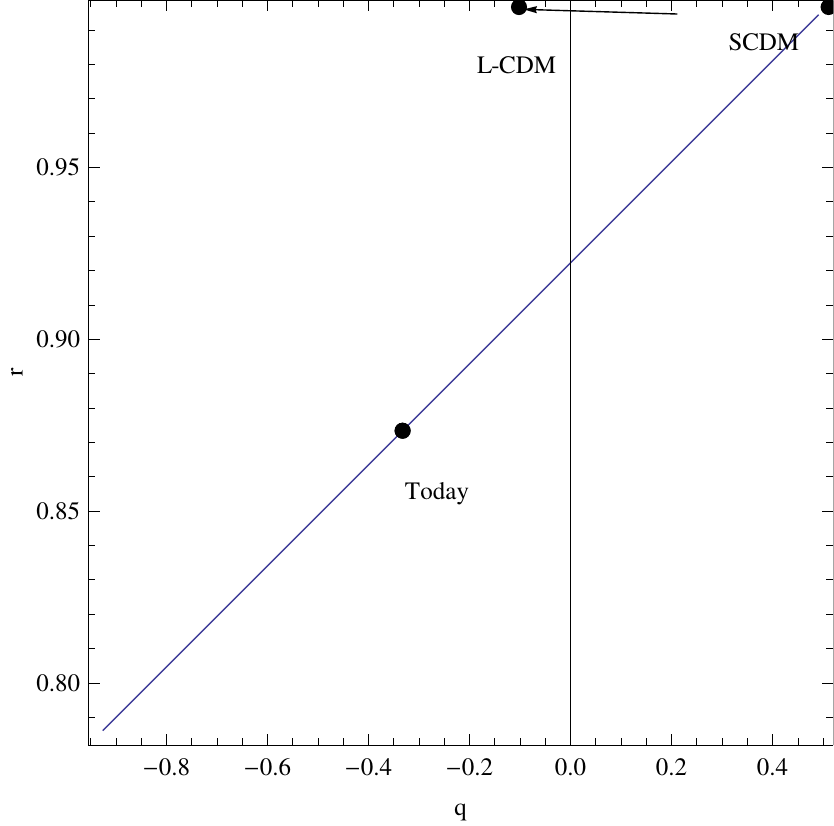}
\includegraphics[scale=0.65]{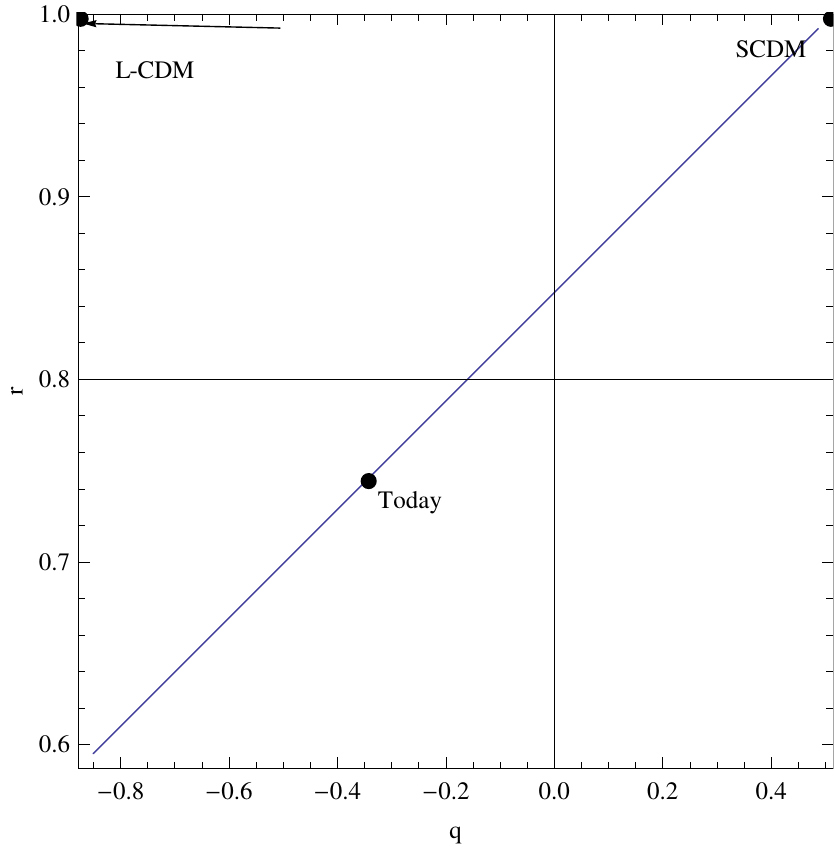}
\includegraphics[scale=0.65]{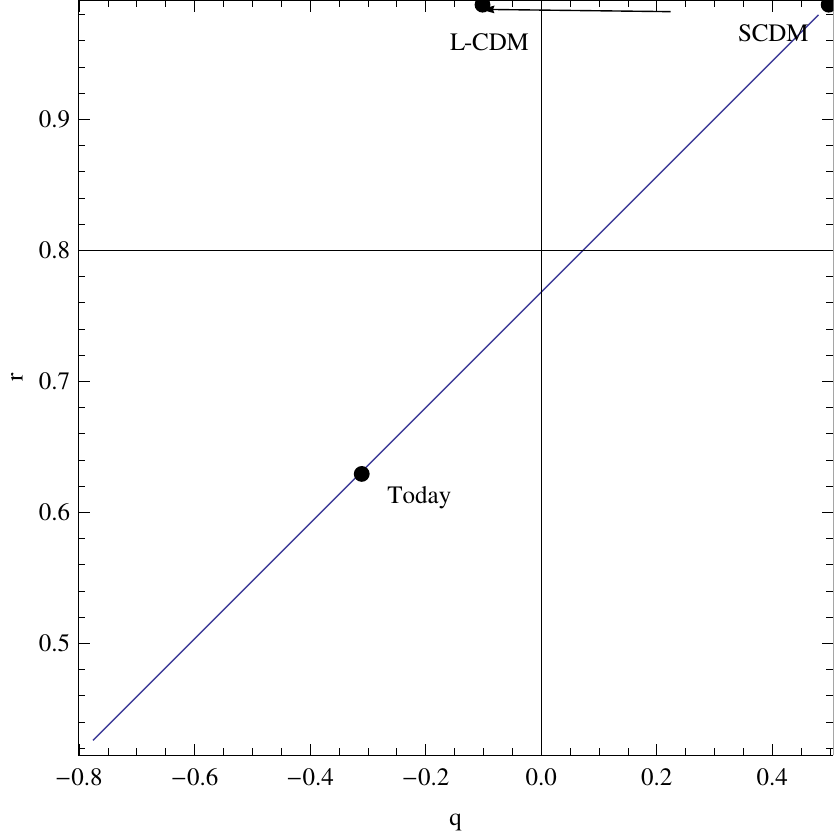}
\caption{Evolution of the interactive MHRDE model in the $r-q$ plane for parameters $(\alpha, \beta)$=(1.33, 0.05), (1.2, 0.1), 
(1.15,0.15) with 
b=0.001} 
\label{fig:rq4}
\end{figure}

In figures \ref{fig:rq3} and \ref{fig:rq4} we have plotted the evolution of the IMHRDE2 model in the $r-q$ plane. The time evolution in 
the $r-q$ plane is from right to left in the plot.
and
For negative values of $\beta$, as seen in figure \ref{fig:rq3}, both IMHRDE2 and $\Lambda$CDM  models commence evolving from the same 
point in the past, corresponds to $r$=1, $q$=0.5, which corresponds to the matter dominates SCDM universe. The $\Lambda$CDM models ends 
with $r$=1, $q$=-1 corresponds to the de Sitter phase, while the IMHRDE2 model, evolves in a different way. For positive 
values of $\beta$, the $r-q$ behavior is as shown in figure \ref{fig:rq4}, where the behavior is almost of the same characteristics. The 
present position of the universe in the $r-q$ plane for negative $\beta$ values are $(r_0,q_0)$=(1.14,-0.460 corresponds to parameters 
(4/3,-0.05), (1.31,-0.57) corresponds to (1.2,-0.1) and (1.03,-0.55) corresponds to (1.01,-0.01). For positive $\beta$ values 
the present position are (0.87,-0.35) corresponds to (4/3,0.05), (0.74,-0.36) corresponds (1.2,0.1) and (0.62,-0.3) corresponds to 
(1.15,0.15). By comparing the present deceleration parameter corresponds to different parameter sets with the observationally 
constraint value, $q_0=-0.60$, the parameter sets (1.01,-0.01) and (1.2,-0.1) are found to be good.  
Out of these two, the parameters (1.2,-0.1) leads to phantom behavior as explained 
earlier so the present value of the $q-$ parameter by the IMHRDE2 model is taken as -0.55 corresponds to parameters (1.01,-0.01). 
It is also seen 
from \ref{fig:rq3} that corresponds to the parameters (1.01,-0.01) the evolution trajectories of IMHRDE1 and $\Lambda$CDM are close to 
each other compared other parameters.

\subsection{Interacting model with $Q=3bH \rho_{de}$-IMHRDE3}

\begin{figure}[h]
 \includegraphics[scale=0.75]{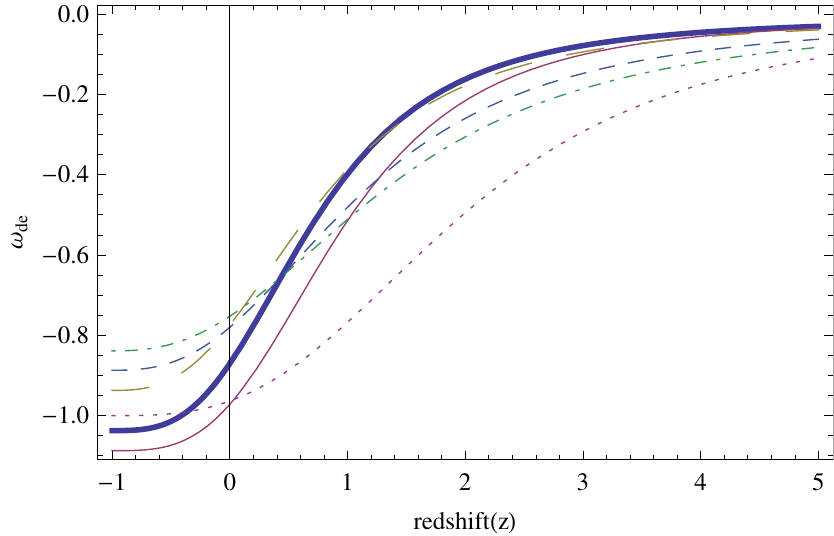}
\caption{Evolution of the equation of state parameter $b$=0.009, for parameters $(\alpha, \beta)$=(4/3,-0.05)(thick continuous line),
(4/3,0.05)(small dashed line),(1.2,-0.1)(thin continuous line),(1.2,0.1)(large dashed line), (1.01,-0.01) (doted line)
 and (1.15,0.15)(dash-dot line).}
\label{fig:eos3}
\end{figure}

In this section we consider IMHRDE model, with $Q=3bH\rho_{de}.$ The second order differential equation in $h^2$ is found to be,
\begin{equation} \label{eqn:diff3}
 {d^2h^2 \over dx^2} + 3(\beta +b+1) {dh^2 \over dx} + 9 (\alpha b + \beta) h^2 =0
\end{equation}
This can be solved as,
\begin{equation} \label{eqn:h3}
 h^2 = f_1 e^{\frac{u_1}{2} x} + f_2 e^{\frac{u_2}{2} x}
\end{equation}
where
\begin{equation}
 f_1={ 3-6\alpha + 3b+3\beta-\sqrt{-36(\alpha b + \beta)+9 (1+b+\beta)^2} + 6(\alpha - \beta)\Omega_{de0} \over 
2b-2\sqrt{-36(\alpha b + \beta)+9 (1+b+\beta)^2} }, \, \, \, \, \, \, \, f_2 = 1-f_1
\end{equation}
and
\begin{equation}
 u_1=-3-b-3\beta-\sqrt{9(1+b+\beta)^2-36(\alpha b+\beta)}, \, \, \, \, u_2=-3-3b-3\beta+\sqrt{9(1+b+\beta)^2-36(\alpha b+\beta)}
\end{equation}
The density parameter and the equation of state in this case is obtained as,
\begin{equation}
 \Omega_{de}= f_1 e^{\frac{u_1}{2} x} + f_2 e^{\frac{u_2}{2} x} - \Omega_{m0} e^{-3x}
\end{equation}
and
\begin{equation}
 \omega_{de}=-1-\left[ {f_1 \frac{u_1}{2} e^{\frac{u_1}{2}x} + f_2 \frac{u_2}{2} e^{\frac{u_2}{2} x} + 3\Omega_{m0} e^{-3x} \over 
3 \left(f_1 e^{\frac{u_1}{2} x} + f_2 e^{\frac{u_2}{2} x} - \Omega_{m0} e^{-3x}  \right)} \right]
\end{equation}
For non-interacting case with $b$=0, and by taking $\Omega_{de0}$=1, the constant coefficients become, $f_1$=1, $f_2$=0, $u_1$=-6$\beta$ 
and $u_2$=-6, consequently the equation of state parameter reduces to 
$\omega_{de}=-1 +\beta$ \cite{tkm1}. The value of the coupling constant $b$ is chosen to be $b=0.009$, up to which, this model is 
compatible with the co-evolution of dark energy and dark matter. For values $b>0.009$ the model fails to explain the coincidence 
problem. With this value of $b$, the evolution of the equation of state parameter
is as shown figure \ref{fig:eos3}
In this model also, the equation of state parameter of IMHRDE3 is starting form zero in the past stage of universe and evolving to negative 
values as the universe expands. In the far future, $\omega_{de}$ approaches to value less than -1 for parameters (4/3,-0.05) and 
(1.2,-0.1). So for these parameters the IMHRDE3 leads to phantom behavior in the far future \cite{Li4}. But for parameters (1.01,-0.01) 
the equation of state approaches -1 as $z\rightarrow -1.$  While for positive $\beta$, the equation of state parameter saturates at 
values greater than -1 in the far future of the universe.  The present values of the equation of state parameter for negative values 
of $\beta$, are -0.86 corresponds to the parameters (4/3,-0.05), -0.97 corresponds to (1.2,-0.1) and -0.97 corresponds to (1.01,-0.01).
For positive values of $\beta$ the value $\omega_{de0}$ are -0.78 corresponds to (4/3,0.05), -0.78 corresponds to (1.2,-0.1) and -0.75 
corresponds to (1.15,-0.15). The best fit is found to be -0.97 corresponds to the parameter $(\alpha,\beta)$=(1.01,-0.01), which is
in confirmation with the corresponding WMAP value $\omega_{de0}$=-0.93.

\begin{figure}[h]
\includegraphics[scale=0.6]{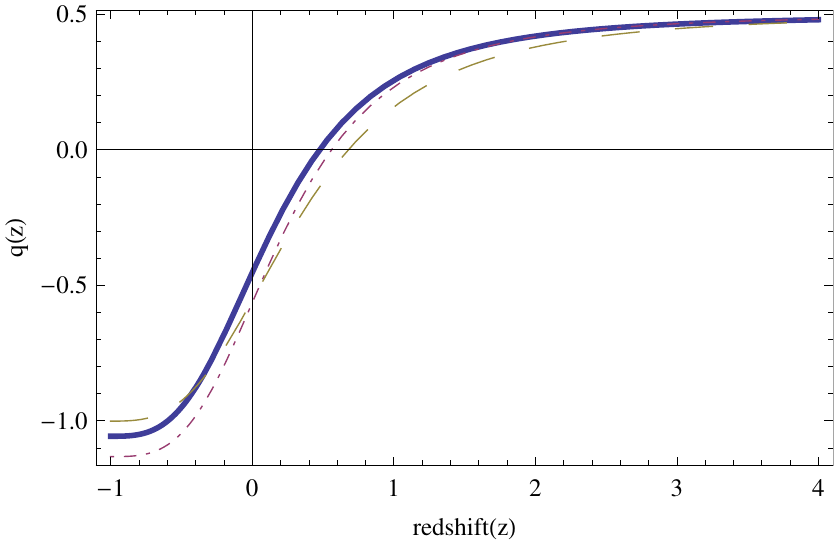}
\includegraphics[scale=0.6]{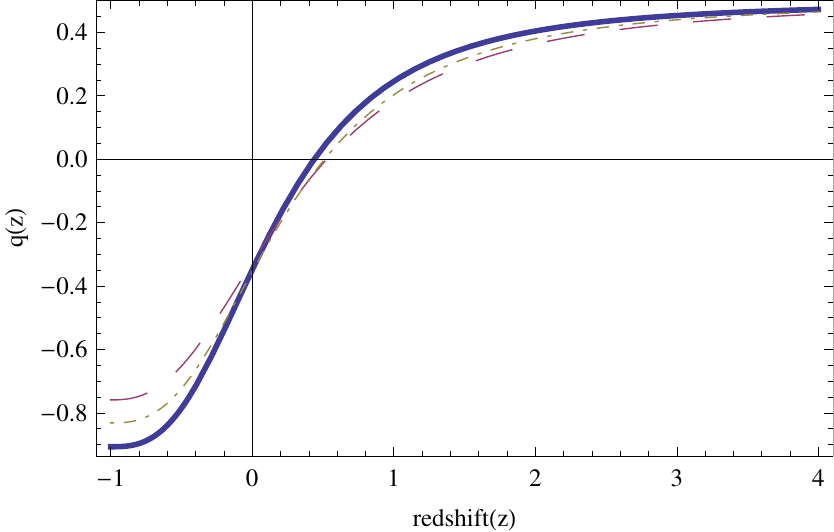}
\caption{Evolution of the $q$-parameter with redshift. Left panel is for parameters $(\alpha,\beta)$=(4/3,-0.05)(thick continuous line),
(1.2,-0.1)(dot-dashed line) and (1.01,-0.01)(dashed line). Right panel is for parameters (4/3,0.05)(thick continuous line), (1.15,0.15)(dot-dashed) and 
(1.2,0.1)(dashed line) with $b$=0.009.}
\label{fig:q3}
\end{figure}

Next we analyses the evolution of the deceleration parameter for this case. With the solution (\ref{eqn:h3}), the deceleration 
parameter can be obtained as,
\begin{equation}
 q=-1 - \frac{1}{2} \left[ { \frac{u_1}{2}f_1 e^{\frac{u_1}{2} x} + \frac{u_2}{2}f_2 e^{\frac{u_2}{2} x} \over 
f_1 e^{\frac{u_1}{2} x} + f_2 e^{\frac{u_2}{2} x} } \right]
\end{equation}
For the non-interacting limit and with $\Omega_{de}$=1, the deceleration parameter reduces to $q=(3\beta -2)/2,$ which is in confirmation 
with the earlier results \cite{tkm1}. The evolution of the $q$ parameter as the universe expands is shown in the figure \ref{fig:q3}.
 In this model also it is clear from figure \ref{fig:q3} that, the deceleration parameter, is starting with 0.5 in the remote past
 of the universe. For negative 
values of $\beta$, the $q$ parameter approaches values less than -1 in the far future for (4/3,-0.05) and (1.2,-0.1). But for parameters 
(1.01,-0.01) the deceleration parameter $q$  approaches -1 as $z\rightarrow -1.$ For positive values of $\beta$, the $q-$parameter
approaches values less than -1 for the three sets of parameters we have used. The present value for the deceleration parameter 
for different $(\alpha,\beta)$  are  -0.47 corresponds to (4/3,-0.05), -0.57 corresponds to (1.2,-0.1),-0.57 corresponds to (1.01,-0.01),
-0.36 corresponds to (4/3,0.05), -0.36 corresponds to (1.2,0.1) and -0.34 corresponds to (1.15,0.15). With the constraints 
from the observational data and also by the fact that the future universe doesn't show any phantom behavior, we are concluding that the best 
fit value for $q_0$ is -0.57 corresponds to the parameters (1.01, -0.01).

From the figure \ref{fig:q3}, is seen that the redshift at which the universe entering the accelerating phase is $z_T$=0.49 
corresponds to (4/3,-0.05), 0.60 corresponds to (1.2,-0.1), 0.70 corresponds to (1.01,-0,01), 0.45 corresponds to (4/3, 0.05), 0.50 
corresponds to (1.2,0.1) and 0.53 corresponds to (1.15,0.15). The observational data prediction for the transition redshift is  
0.45 - 0.73 \cite{Komatsu1}. In the light of this the best fit value from this model is $z_T$=0.70 corresponds to 
$(\alpha,\beta)$=(1.01,-0.01).

\subsubsection{Statefinder analysis}

The statefinder parameters $(r,s)$ can be obtained for IMHRDE3, by using the solution (\ref{eqn:h3}) as,
\begin{equation}
 r=1+ \left[ { \frac{u_1^2}{4}f_1e^{\frac{u_1}{2} x} + \frac{u_2^2}{4}f_2e^{\frac{u_2}{2}x}+\frac{3}{2}u_1f_1e^{\frac{u_1}{2}x} +
\frac{3}{2}u_2f_2e^{\frac{u_2}{2}x} \over 2 (f_1 e^{\frac{u_1}{2}x} + f_2e^{\frac{u_2}{2}x})} \right]
\end{equation}
and
\begin{equation}
 s=-\left[ { \frac{u_1^2}{4}f_1e^{\frac{u_1}{2} x} + \frac{u_2^2}{4}f_2e^{\frac{u_2}{2}x}+\frac{3}{2}u_1f_1e^{\frac{u_1}{2}x} +
\frac{3}{2}u_2f_2e^{\frac{u_2}{2}x} \over \frac{3}{2}u_1f_1e^{\frac{u_1}{2}x}+\frac{3}{2}u_2f_2e^{\frac{u_2}{2}x}+9f_1e^{\frac{u_1}{2}x}
+9f_2e^{\frac{u-2}{2}x} } \right]
\end{equation}
In non-interacting limit, $b$=0 with $\Omega_{de}$=1, the statefinder parameters reduces to the standard form $r=1+9\beta(\beta -1)/2$ 
and $s=\beta.$ The $r-s$ plots for the present model is shown in the figures below.
The figures \ref{fig:rs31} and \ref{fig:rs32} shows the behavior of IMHRDE3 in the $r-s$ plane. The evolution of the $r-s$ is from 
right to the left as in two previous cases. The today's position of the universe and the $\Lambda$CDM fixed point (LCDM point) are noted 
in all the figures. It is clear that the distance between the present position of the universe and the $\Lambda$CDM is decreasing as 
$\beta$ increases for negative values of $\beta.$ While for positive values of $\beta$ the separation is increasing the $r-s$ plane. Also the 
$\Lambda$CDM point is lying on the $r-s$ plane for negative values of $\beta.$ But for the positive $\beta$ values, the $\Lambda$CDM point is out of 
the $r-s$ trajectory. An important point to be noted regarding the $r-s$ behavior corresponds to the parameters (1.01, -0.01), for 
which the $\Lambda$CDM point is lying on the future phase of the evolution of the universe. This indicating that for the best parameters, 
(1.01,-0.01) the universe is tending towards the $\lambda$CDM phase in the future. On the contrary for other parameters the 
$\Lambda$CDM point 
is lying on the past phase of the $r-s$ plots as it is clear from figure \ref{fig:rs31}.
\begin{figure}[h]
 \includegraphics[scale=0.5]{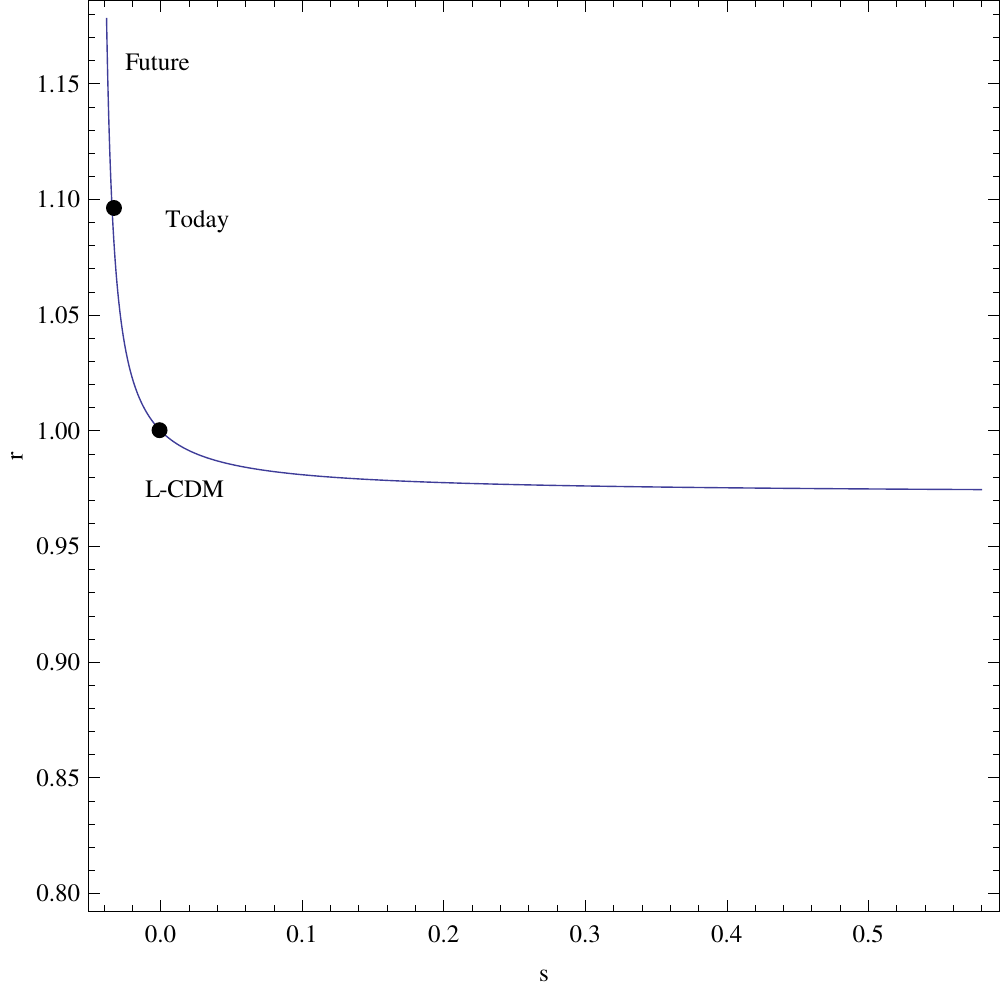}
\includegraphics[scale=0.5]{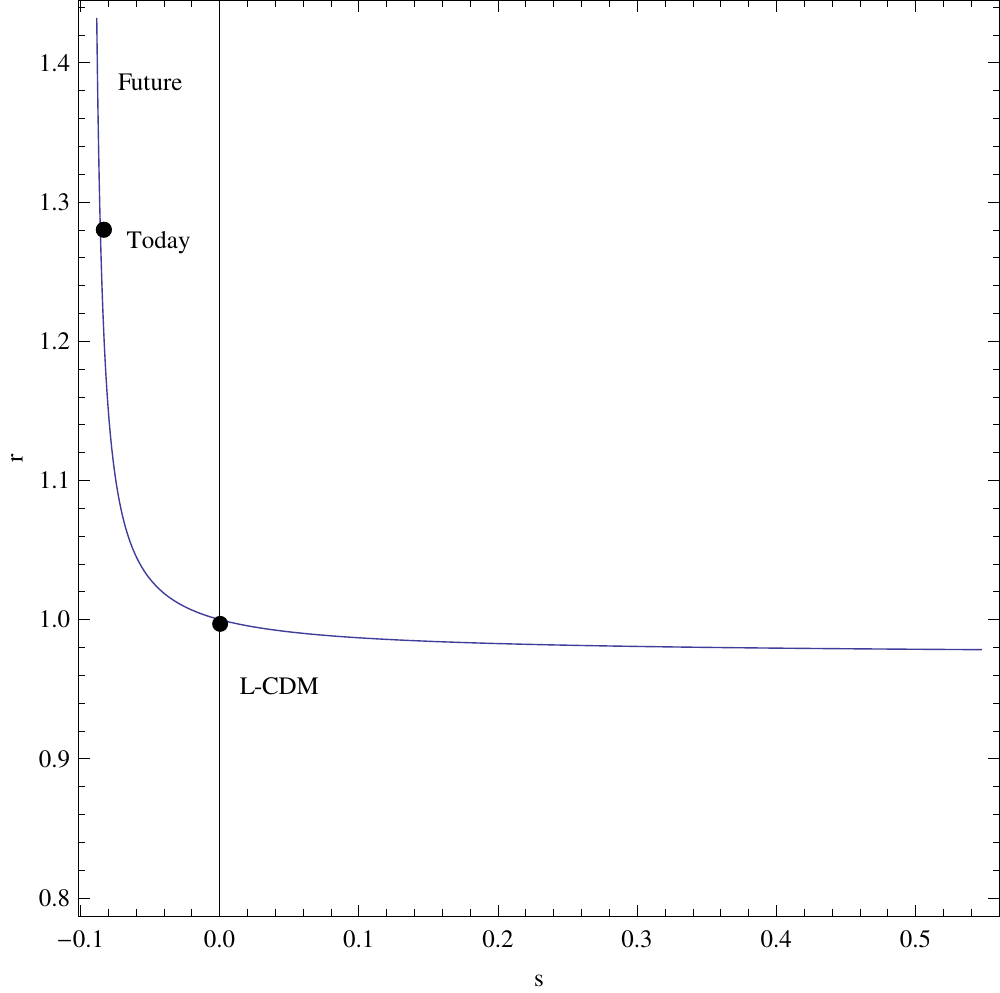}
\includegraphics[scale=0.5]{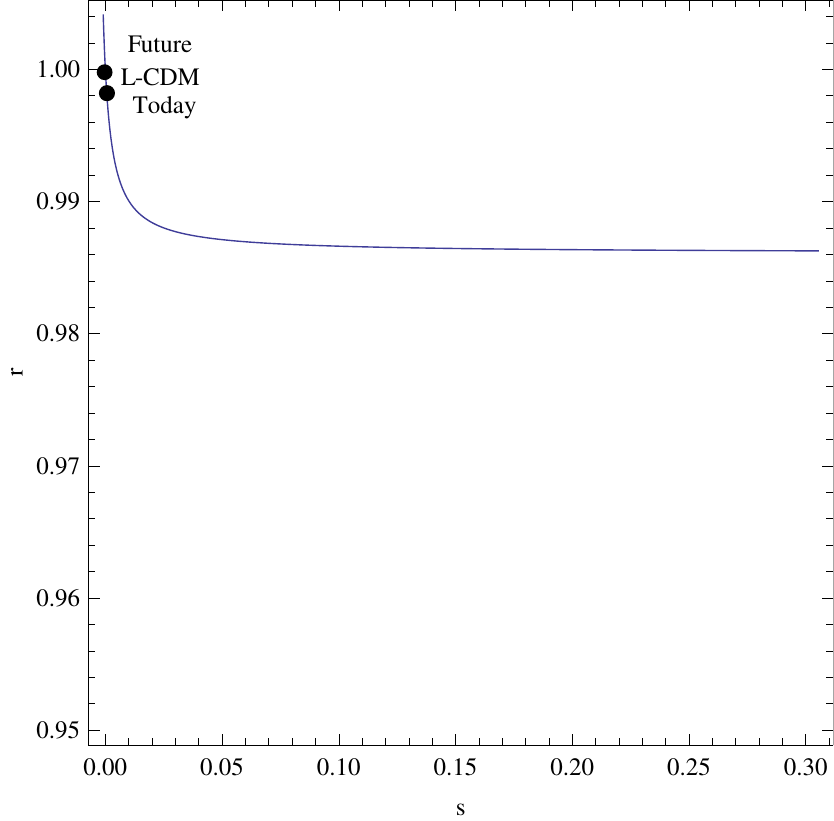}
\caption{$r-s$ plots for parameters $(\alpha,\beta)$=(4/3,-0.05)(left panel), (1.2,-0.1)(middle panel), (1.01,-0.01) 
(the right panel) with $b$=0.009}
\label{fig:rs31}
\end{figure}

\begin{figure}[h]
 \includegraphics[scale=0.5]{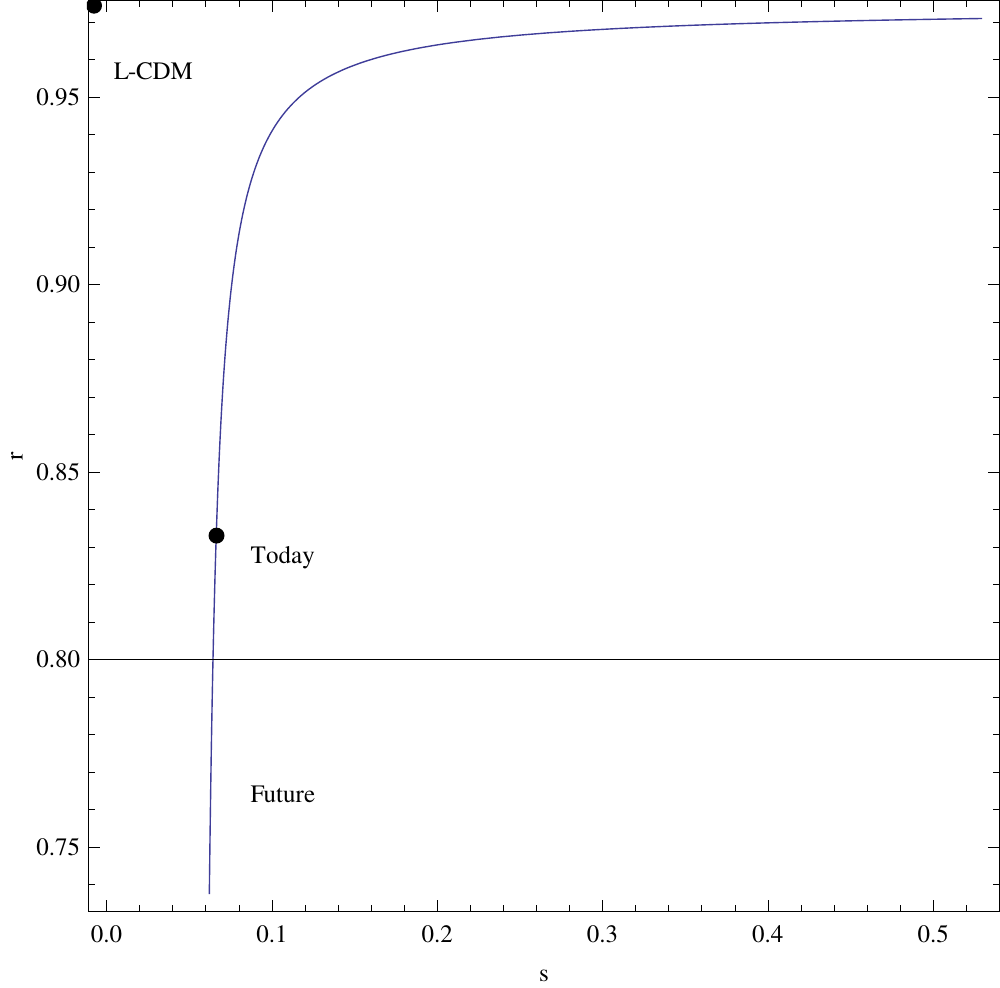}
\includegraphics[scale=0.5]{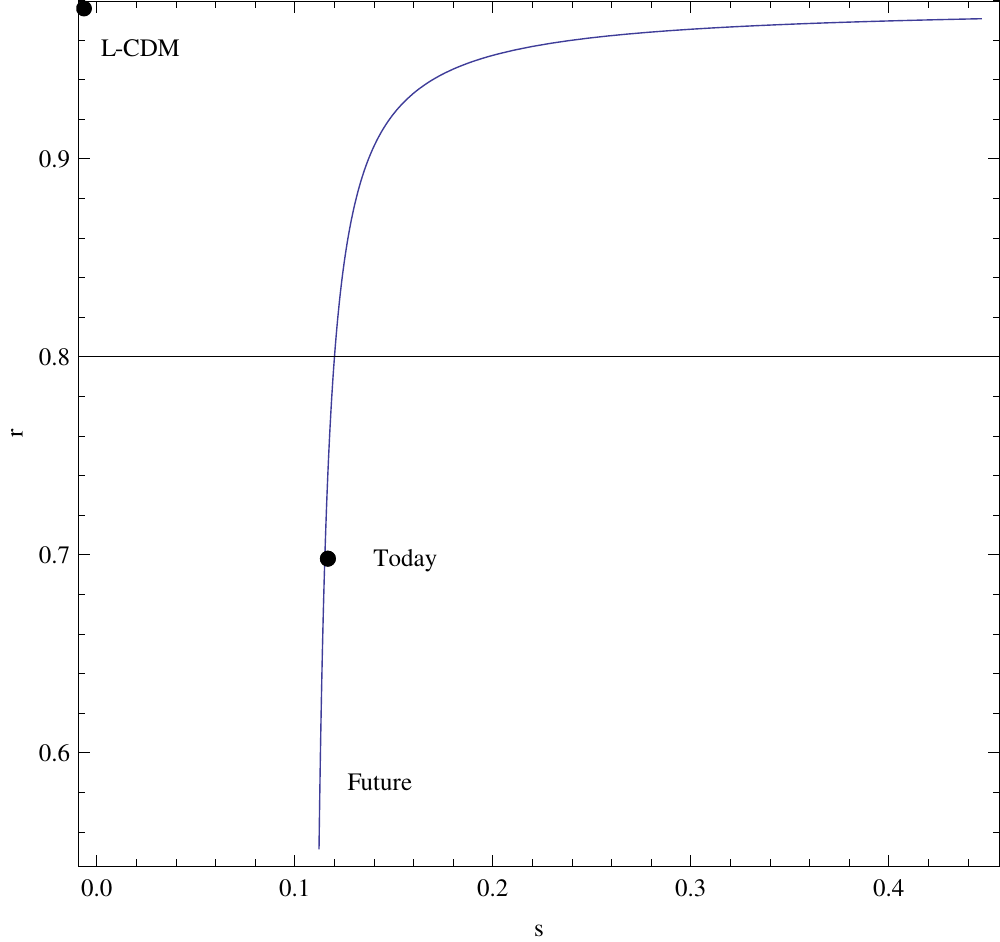}
\includegraphics[scale=0.5]{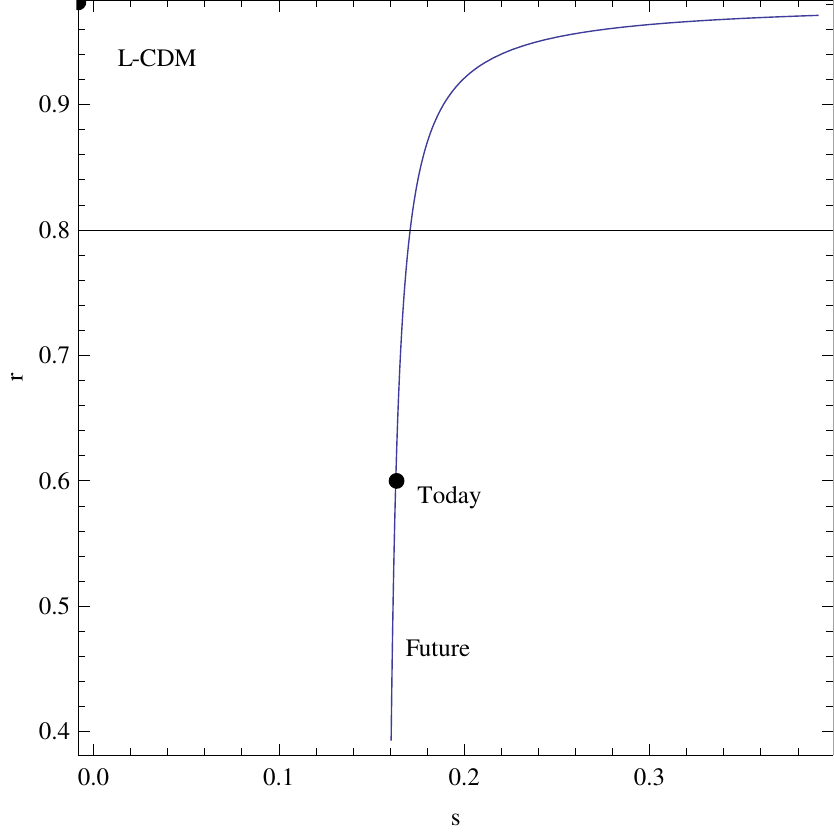}
\caption{$r-s$ plots for parameters $(\alpha,\beta)$=(4/3,0.05)(left panel), (1.2,0.1)(middle panel), and (1.15,0.15)(right panel)
 with $b$=0.009}
\label{fig:rs32}
\end{figure}

The present position of the universe in the $r-s$ plots are $(r_0,s_0)$=(1.1,-0.034) corresponds to (4/3,-0.05), (1.27,-0.085) 
corresponds to (1.2,-0.1), (0.98,0.0004) corresponds to (1.01,-0.01), (0.83,0.066) corresponds to (4/3,0.05), (0.70,0.115) corresponds 
to (1.2,0.1) and (0.60,0.16) corresponds to (1.15,0.15). For the best fit parameters, (1.01,-0.01) the present position in the 
$r-s$ plane is distinguishing the IMHRDE3 from other standard models. For the new HDE model \cite{Setare1}, the present 
position of the universe is corresponds to $(r_0,s_0)$=(1.357,-0.102). Compared to the new HDE model, the present IMHRDE3 model is
comparatively much closer to the $\Lambda$CDM model.

\begin{figure}[h]
\includegraphics[scale=0.5]{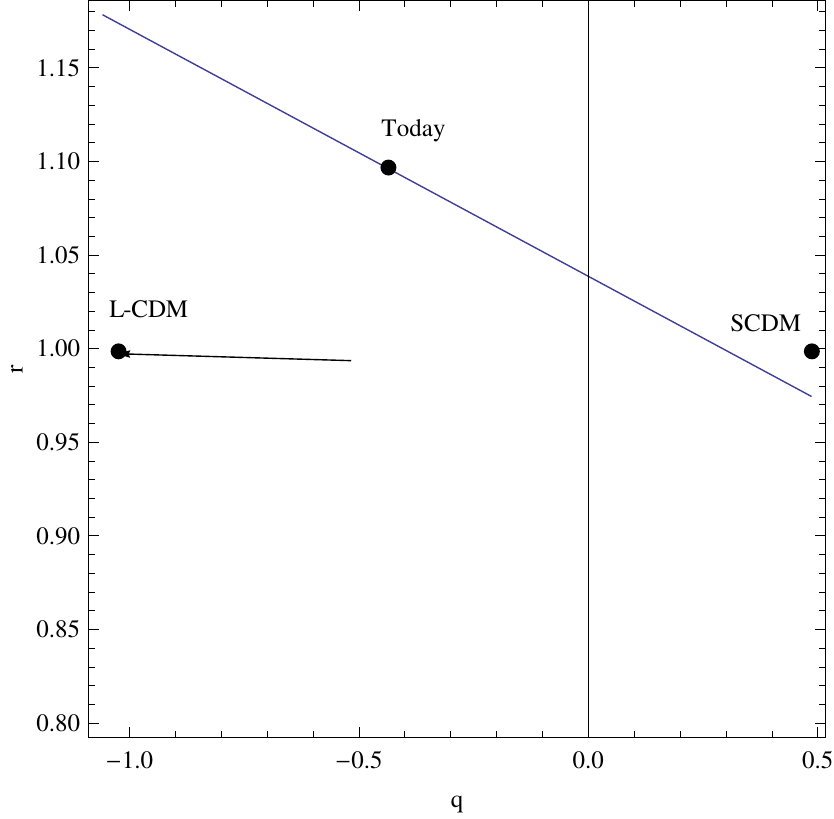}
\includegraphics[scale=0.5]{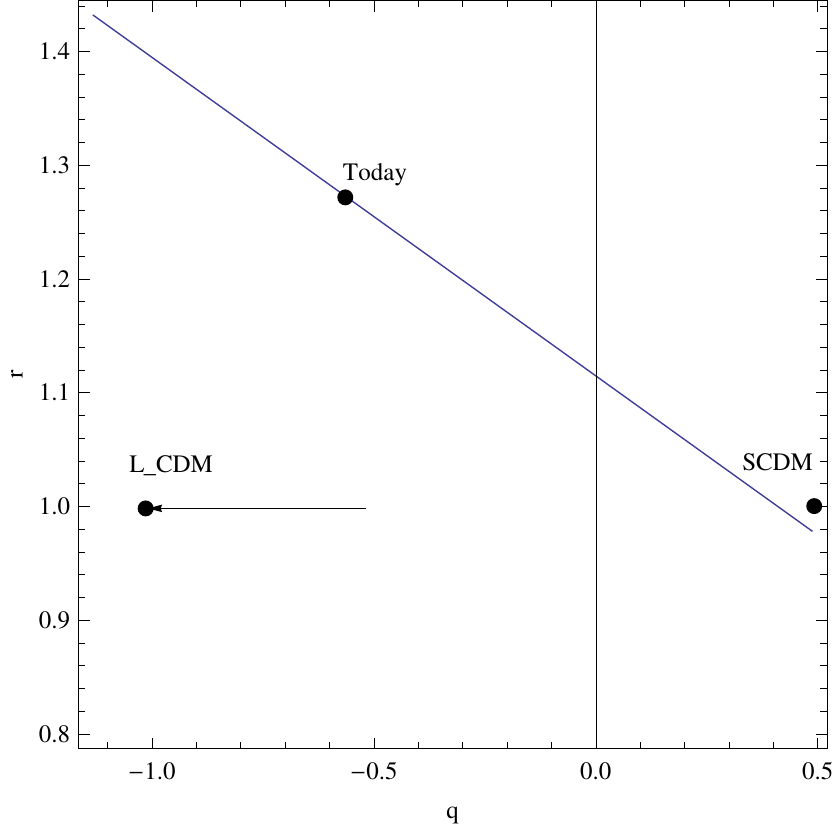}
\includegraphics[scale=0.5]{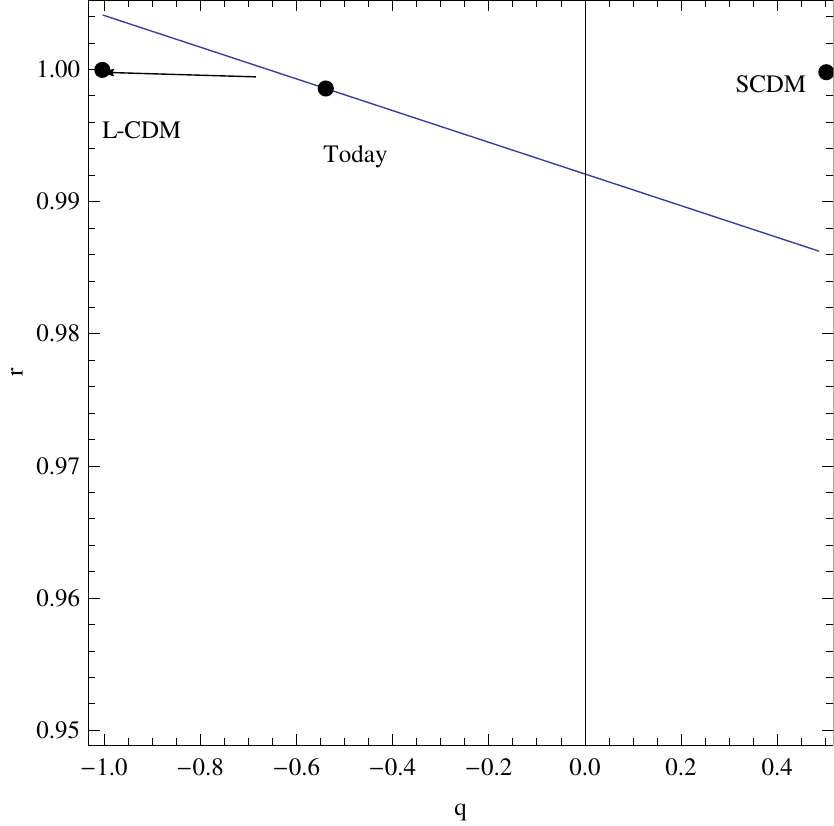}
\caption{$r-q$ behavior of the IMHRDE for parameters $(\alpha, \beta)$=(4/3,-0.05)(right figure,(1.2,-0.1)(middle) and (1.01,-0.10)
(right), with $b$=0.009}
 \label{fig:rq5}
\end{figure}

\begin{figure}[h]
\includegraphics[scale=0.5]{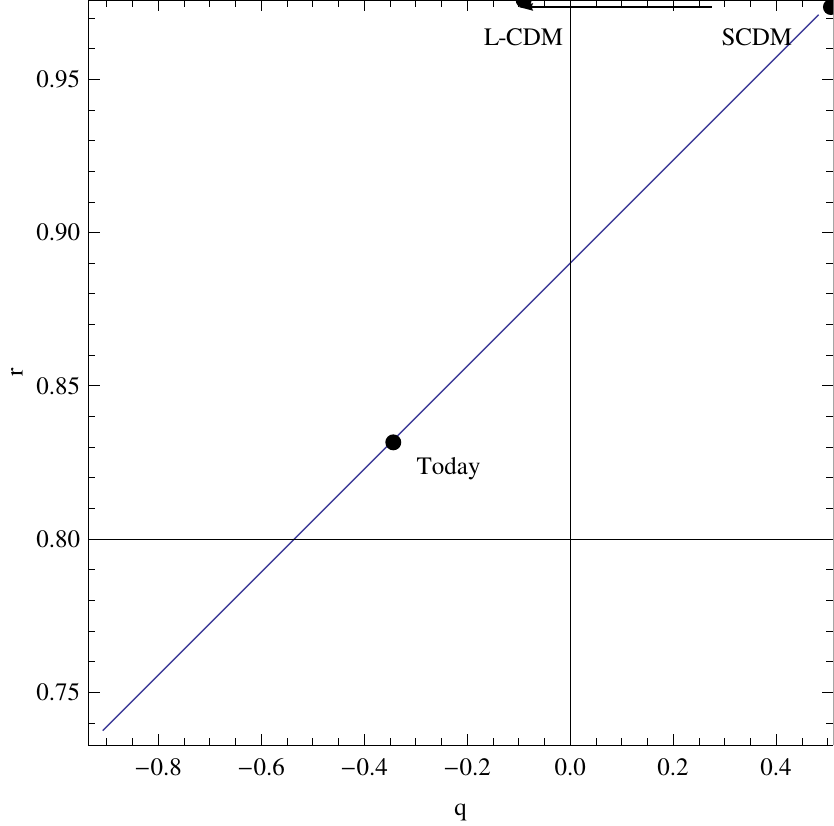}
\includegraphics[scale=0.5]{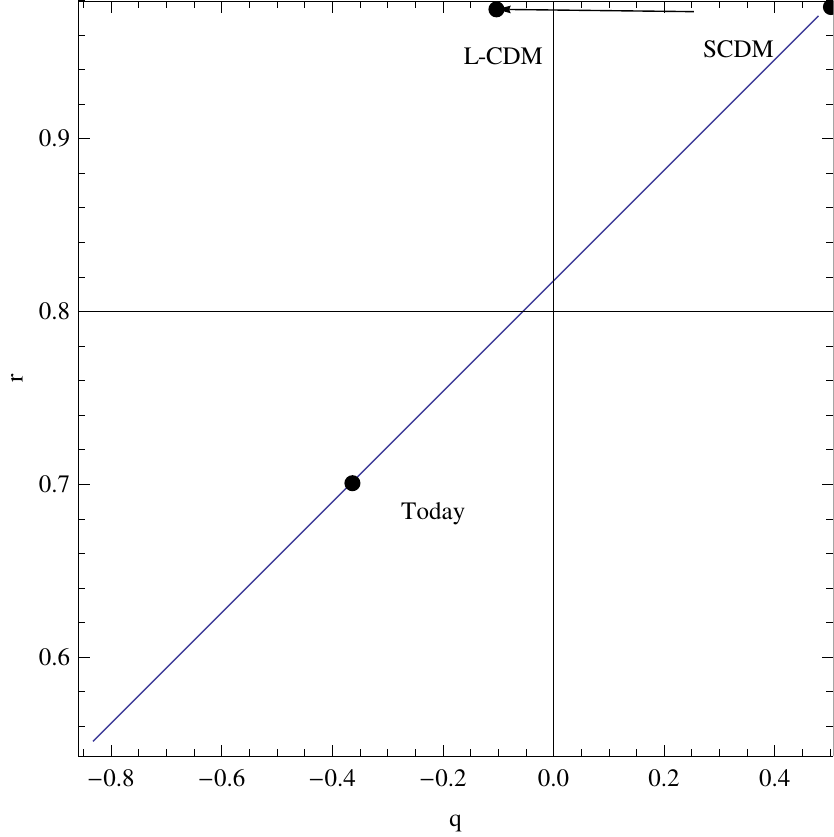}
\includegraphics[scale=0.5]{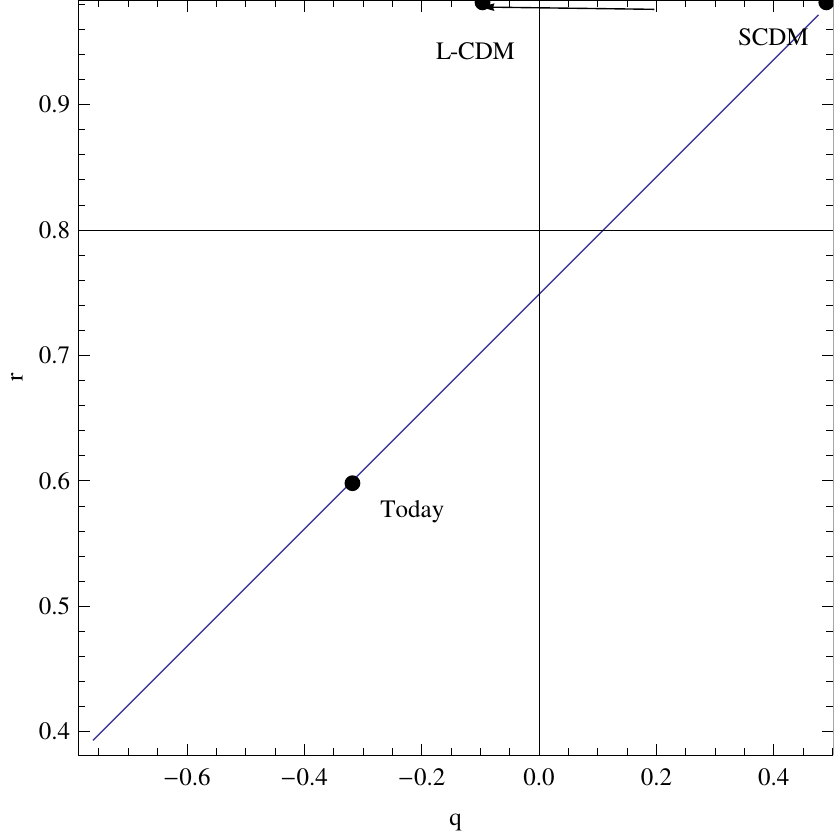}
\caption{$r-q$ behavior of the IMHRDE for parameters $(\alpha, \beta)$=(4/3,0.05)(left plot),(1.2,0.1)(middle plot) and (1.15,0.15)
(right plot), with $b$=0.009}
 \label{fig:rq6}
\end{figure}

In order to confirm the $r-s$ behavior of IMHRDE3, we analyses 
the behavior of this model in the $r-q$ plane also. The respective plots are given in figures \ref{fig:rq5} and \ref{fig:rq6}.

The evolution in the $r-q$ plane is from right to left. For negative values of $\beta$, none of the $r-q$ plots are starting from the 
SCDM phase, although for parameters (1.2,-0.1), the starting point in the $r-q$ plane is very much close to the SCDM point. 
For the negative $\beta$ values we have considered, the distance of the starting point of IMHRDE3 and SCDM point in $r-q$ plane is the largest 
for the parameters $(1.01,-0.01)$, which is our best fit parameters. Also to be noted that the distance between the $\lambda$CDM point 
and today's position in the $r-q$ plane is the smallest for the parameters (1.01,-0.01). For positive values of $\beta$ all the $r-q$ 
plots commences from the SCDM phase and the distance of the today's position and $\Lambda$CDM point is increasing as $\beta$ increases.
The present value of the $r-q$ parameters are (1.1,-0.45) corresponds to (4/3,-0.05), (1.27,-0.57) corresponds to (1.2,-0.1),
(0.98,-0.57) corresponds to (1.01,-0.01), (0.83,-0.35) corresponds to (4/3,0.05), (0.70,-0.360 corresponds to (1.2,0.1) and 
(0.60,-0.33) corresponds to (1.15,0.15). For the best parameters (1.01,-0.01), the present parameters $(r_0,q_0)$=(0.98,-0.57), 
for which 
the $q_0$ value is very close the WMAP value -0.60. So for IMHRDE3 also (1.01,-0.01) is the best fit parameters corresponds to which,
the predicted cosmological parameters are $\omega_{de0}$=-0.97, $z_T$=0.70, $q_0$=-0.57.

\section{Conclusions}

We have considered the modified holographic Ricci dark energy interacting with the dark matter in a flat universe. The interaction is 
non-gravitational and linear. We consider three phenomenological form for the interaction term $Q$, which is basically proportional to the 
Hubble parameter $H$ and the densities of dark energy and dark matter. The three interaction forms are $Q=3bH(\rho_{de}+\rho_m)$
(IMHRDE1), $3bH\rho_m$ (IMHRDE2),$ 3bH\rho_{de}$ (IMHRDE3). We have considered these three interaction cases separately, and studied the 
the evolution of the equation of state, deceleration parameter and also made the statefinder diagnostic analysis to discriminate 
the models form other standard models. 

In the case of IMHRDE1, we have found that the interaction coupling constant can at most be $b=0.001.$, for values higher than 
this the IMHRDE1 is found to be incompatible with coincidence between dark energy and dark matter. So we 
took $b=0.001$ for analyzing IMHRDE1. Further we have found that the best fit parameters are $(\alpha,\beta)$=(1.01,-0.01). 
The corresponding the equation of 
state parameter evolves in such a way that at far future in the evolution of the universe , as $z\rightarrow -1$, 
$\omega_{de} \rightarrow -1.$ This implies that the IMHRDE1 
approaching a de Sitter phase in the far future. The present value of the equation of state parameter is found to be as $\omega_{de}=-0.96$
This value is agreeing closely with the value reported by the WMAP project \cite{Komatsu1}, as -0.93. The transition of the universe to 
accelerating expansion is found to occur at redshift $z_T=0.70$ This value is in agreement with the observational constraint 
$z_T$=0.45 - 0.73 obtained form the analysis of SNe+CMB data with $\lambda$CDM model. We have also obtained the evolution of the 
deceleration parameter $q$ of IMHRDE1 model. For best parameters $(\alpha,\beta)$=(1.01,-0.01), we have found that the deceleration 
parameter approaches -1 as $z \rightarrow -1,$ which again shows that the at future evolution the universe tends to the behavior of 
de Sitter universe. The present value of the deceleration parameter is obtained as $q_0=-0.56.$ The WMAP data constraint the $q_0$ as
-0.60. So the IMHRDE1 value of $q_0$ is very close to this observationally constraint value. In a work by Luis et. al.\cite{Luis1} 
have found that the value of $q_0=-0.59$, for the IMHRDE with non-linear interaction between dark energy and dark matter for the same 
$(\alpha,\beta)$ parameters.

The statefinder analysis of the IMHRDE1 model have shown that for the best parameters, the statefinder parameter have the present 
value as $(r_0,s_0)$=1.03,-0.008. Compared the standard $\lambda$CDM model, with $(r_0,s_0)$=(1,0), so it is clear that the distance 
between the present IMHRDE1 model and $\Lambda$CDM model is very small in the $r-s$ plane. But compared to other standard models, for 
example, the Chaplygin gas model of dark energy, for which $(r>1, s<0)$, the present IMHRDE1 phase is different. However as universe 
evolves $r-s$ behavior of IMHRDE1 (\ref{fig:rs1}) is approaching the Chaplygin gas behavior. But compared to HDE model with event 
horizon as the IR cut-off, for which $(r,s)$=(1, 2/3), the IMHRDE1 is shows a different evolution. These are further verified with 
checking the evolution of the IMHRDE1 in $r-q$ plane.

In IMHRDE2 model the interaction coupling constant if found to be higher as $b=0.003,$ compared to IMHRDE1. For $b$ values 
above this the model is not compatible with the co-evolution of the dark sectors. The equation state evolution is obtained and 
found that for the best fit parameters $(\alpha,\beta)$=(1.01,-0.01), the $\omega_{de} \rightarrow -1$ as redshift $z \rightarrow -1.$
This implies that, in the far future the IMHRDE2 model also tending towards a de Sitter type evolution. The present value of the equation 
of state parameter is around $\omega_{de0}=-0.96,$ which is close to WMAP value -0.93. The evolution of the deceleration parameter 
is shown in figure \ref{fig:q212}. Accordingly the transition redshift is found to $z_T=0.68$ for best fit parameters, which is in close agreement with 
observationally constraint range \cite{Alam1}. The present value of the $q-$parameter of IMHRDE2 is $q_0=-0.55$ for the best fit 
parameters, and is close agreement with the corresponding value predicted WMAP data.

The statefinder evolution is studied for the IMHRDE2, and found that, the evolution is almost similar with that of IMHRDE1. The 
present values of the statefinder parameters is found to be $(r_0,s_0)$=(1.03,-0.0096), which implies that the IMHRDE2 will 
behave as Chaplygin gas in the future. In discriminating the IMHRDE2 from other models, the present position of IMHRDE2 is different 
compared to the $\Lambda$CDM model with (1,0) and HDE model with event horizon as IR cut-off with (1,2/3). The $r-q$ plane plot seen to 
be compatible with the above conclusions. 

For the analysis of IMHRDE3 we have chosen the coupling constant as $b=0.009,$ because for values higher than this, the model 
does not predict the co-evolution of the dark sectors. The analysis of the evolution of equation of state parameter for this have clearly 
shown that $\omega_{de} \rightarrow -1$ as $z \rightarrow -1$ for the best fit parameters (1.01,-0.01). This implies that as like other two models this model also tending 
towards a de Sitter evolution in the far future. The present equation of state parameter for the best fit $(\alpha,\beta)$ is found to 
around $\omega_{de0}$=-0.97, which is close to the observationally constraint value. The deceleration parameter of this model is evolved 
in such that, the transition to accelerating phase is occurred at around $z_T$=0.70, and is evidently very well in the observational range.
The present value of $q_0$ is around -0.57 for best fit parameters, which is quite close to WMAP prediction.

The IMHRDE3 model was discriminated form other standard models using the $r-s$ diagnosis. The behavior of the model in the $r-s$ plane 
have shown that, for the best fit parameters, the model is evolving to the $\Lambda$CDM phase in the future. This shows that 
the de Sitter evolution of the IMHRDE3 in future may end on the $\Lambda$CDM phase. The present statefinder parameters are $(r_0,s_0)$=
(0.98,0.0004), which is close to the $\Lambda$CDM point and different from the HDE model with event horizon as the IR cut-off.

In summary we have considered the IMHRDE model with possible interaction between the dark energy and dark matter. We have found that all 
the three interacting models will approach the de Sitter phase in the later stages of the evolution of the universe for the best 
parameters $(\alpha,\beta)$=(1.01,-0.01). The best fit $(\alpha,\beta)$ parameters have -ve value for $\beta,$ which is advisable 
because the positivity of dark energy density require to take $\beta<0$ \cite{Luis1}. In particular the IMHRDE3 model evolves to $\Lambda$CDM phase in its future evolution. A similar
kind of work was carried out in reference \cite{Fu1}, with another form for Ricci dark energy 
$\displaystyle \rho_{de}=3 \alpha M_{Pl}^2 (\dot{H} + 2 H^2)$ for the same forms for the interaction term Q, where the authors mainly 
concentrated on evaluating the Hubble Parameter and density parameter for dark matter. In our work we have done the evaluation of different 
parameters and analyses the IMHRDE with statefinder diagnosis.


\begin{thebibliography}{}
\bibitem{Perl} Perlmutter S {\it et al.}    Astrophys. J. {\textbf 517}, (1999) 565. 
\bibitem{Ries} Riess A G {\it  et al.}   Astrophys. J.{\textbf 607}, (2004) 665  [SPIRES]. 
\bibitem{wein1} Weinburg S,   Rev. Mod. Phys. {\textbf 61}, (1989) 1  [SPIRES].
\bibitem{sahni1}  Sahni V and Starobinsky A A,   Int. J. Mod. Phys. D {\textbf 9}, 
\bibitem{Cai1} Y F Cai, E N Saridakis, M R Setare and J Q Xia,  Phys. Rept. \textbf{493}, (2010) 1.
\bibitem{Li1} M Li, X D Li, S Wang and Y Wang  Commun. Theor. \textbf{56}, (2011) 525.
\bibitem{Xang1} X Zhang and F Q Wu  Phys. Rev. D \textbf{72}, (2005) 043524.
\bibitem{Huang1} Q G Huang and M Li,  JCAP \textbf{0503}, (2005) 001. 
\bibitem{Xang2} X Zhang, Int. J. Mod. Phys. D \textbf{14}, (2005) 1597. 
\bibitem{Xang3} X Zhang, Phys. Lett. B \textbf{683}, (2010) 81. 
\bibitem{Skind1} L Susskind,  J. Math. Phys. \textbf{36}, (1995) 6377. 
\bibitem{Cohen1} A. G. Cohen, D. B. Kaplan and A. E. Nelson, Phys. Rev. Lett. \textbf{82} (1999) 4971.
\bibitem{Li2} M. Li, Phys. Lett. B \textbf{603} (2004) 1
\bibitem{Hsu1} Hsu S D H, Phys. Lett. B \textbf{594} (2004) 13
\bibitem{Gao1} Gao F Q, Wu X, Cheng and Y G Shen, Phys. Rev. D \textbf{79} (2009) 043511.
\bibitem{Zhang1} Linsen Zhang, Puxun Wu and Hongwei Yu, Eur. Phys. J. C. \textbf{71} (2011) 1588.
\bibitem{Yang1} Yang R -J, Z -H Zhu and Wu F, Int. J. Mod. Phys. A \textbf{26} (2011) 317.
\bibitem{Granda1} Granda l N and Oliveros A, Phys. Lett. B \textbf{671} (2009) 199.
\bibitem{Luis1} Luis P Chimento and Richarte M G, Phys. Rev. D \textbf{84} (2011) 123507.
\bibitem{Teg1} Tegmark M et. al., Phys. Rev. D \textbf{69} (2004) 103501.
\bibitem{Amend1} Amendola l, Phys. Rev. D \textbf{62} (2000) 043511.
\bibitem{Xang4} X Zhang, Mod. Phys. Lett. A \textbf{20} (2005) 2575.
\bibitem{Nijor1} Nijori S D Odintsov, Phys. Rev D \textbf{72} (2005) 023003
\bibitem{He1} He J H, Wang B and Jing Y P, JCAP \textbf{0907} (2009) 030.
\bibitem{Li3} Li Y H and Zhang X, Eur. Phys. J C \textbf{71} (2011) 1700.
\bibitem{Sahni2} Sahni V, Saini T D, Starobinsky A A and Alam U, JETP Lett. \textbf{77} (2003) 201.
\bibitem{Setare1} Setare M R, Zhang J and Xing Zhang, JCAP \textbf{03} (2007) 007
\bibitem{Malek1} Malekjani M, Khodam-Mohammadi A and Nazari-pooya N, (2010) [arXiv:1011.4805]
\bibitem{Komatsu1} Komatsu E et. al., [WMAP Collaboration], Astrophys. J. Suppl. \textbf{192} (2011) 18.
\bibitem{tkm1} Titus K Mathew, Jishnu Suresh and Divya Divakaran, Int. J. Mod. Phys. D, \textbf{22} (2013) 1350056.
\bibitem{Luis2} Luis P Chimento, Monica Forte and Martin G Richarte, Eur. Phys. J. C \textbf{73} (2013) 10052.
\bibitem{Fu1} Tian-Fu, Jing-Fei Zhang, Jin-Qian Chen and Xing Zhang, Eur. Phys. J C \textbf{72} (2012) 1932.
\bibitem{Chatto1} Chattopadhyay S, Eur. Phys. J Plus \textbf{126} (2011) 130.
\bibitem{Cald1} Caldwell R R, Phys. Lett. B \textbf{545} (2002) 23.
\bibitem{Alam1} Alam U, Sahni V and A A Starobinsky, JCAP \textbf{406} (2004) 008.
\bibitem{Zim1} Zimdahl W and Pavon D, Gen. Rel. Grav. \textbf{36} (2004) 1483.
\bibitem{Li4} Chimento L P, Forte M, Lazkoz R and Richarte M G, Phys. Rev. D \textbf{79} (2009) 043502.
\bibitem{Fu1} Tian-Fu Fu, Jing-Fei Zhang, Jin-Qian Chen and Xin Zhang, Eur. Phys. J C \textbf{72} (2012) 1932.     .
%

\end{thebibliography}
\end{document}